\documentclass[preprint2,tighten]{aastex6} 
\pdfoutput=1 
\usepackage{amsmath,amstext}
\usepackage{apjfonts}
\usepackage[T1]{fontenc}
\usepackage[figure,figure*]{hypcap} 
\usepackage{enumitem}

\usepackage{xcolor}
\definecolor{xlinkcolor}{cmyk}{1,1,0,0}
\hypersetup{
     linkcolor=xlinkcolor,     
     citecolor=xlinkcolor,     
     filecolor=xlinkcolor,  
     urlcolor=xlinkcolor,      
     }
\usepackage{multirow}
\newcommand{\be}{\begin{equation}}
\newcommand{\ee}{\end{equation}}
\newcommand{\bee}{\begin{eqnarray}}
\newcommand{\eee}{\end{eqnarray}}

\newcommand{\avg}[1]{\left\langle{#1}\right\rangle}

\iftrack
\renewcommand{\added}[1]{{\bf #1}}
\renewcommand{\deleted}[1]{\ }
\renewcommand{\replaced}[2]{{\bf #2}}
\fi

\begin{document}
\title{On estimation of contamination from hydrogen cyanide in carbon monoxide\\ line-intensity mapping}
\shorttitle{Hydrogen cyanide in carbon monoxide mapping}
\shortauthors{Chung et al.}
\author{Dongwoo T. Chung\altaffilmark{1}}
\author{Tony Y. Li\altaffilmark{1}}
\author{Marco P. Viero\altaffilmark{1}}
\author{Sarah E. Church\altaffilmark{1}}
\and\author{Risa H. Wechsler\altaffilmark{1,2}}
\altaffiltext{1}{Kavli Institute for Particle Astrophysics and Cosmology; 
  Physics Department, Stanford University, Stanford, CA 94305, USA; \href{mailto:dongwooc@stanford.edu}{dongwooc@stanford.edu}}
\altaffiltext{2}{SLAC National Accelerator Laboratory, Menlo Park, CA 94025, USA}

\begin{abstract}

  Line-intensity mapping surveys probe large-scale structure through spatial variations in molecular line emission from a population of unresolved cosmological sources. Future such surveys of carbon monoxide line emission\replaced{ (}{, }specifically the CO(1-0) line\replaced{)}{,} face potential contamination from a disjoint population of sources emitting in a hydrogen cyanide emission line, HCN(1-0). This paper explores the potential range of the strength of HCN emission and its effect on the CO auto power spectrum, using simulations with an empirical model of the CO/HCN--halo connection.  We find that effects on the observed CO power spectrum \replaced{vary with modeling approaches}{depend on modeling assumptions} but are very small for our fiducial model, \replaced{with}{which is based on current understanding of the galaxy--halo connection. Given the fiducial model, we expect} the \replaced{undesirable boost}{bias} in overall CO detection significance due to HCN\deleted{ expected }to be less than 1\%.
\end{abstract}


\section{Introduction}
\label{sec:intro}
{Line-intensity mapping} is a novel method of mapping large-scale structure in the early universe through space and time, in which broad spatial variations in line-intensity throughout the surveyed volume trace galaxies too faint to be resolved individually in a conventional galaxy survey. Such observations allow for a statistical understanding of a large population of galaxies, without resolving every galaxy individually. The spectral lines that these surveys target arise from certain species of molecular gas abundant in these galaxies, generally connected to star-formation activity. Therefore, line-intensity mapping can contribute to our understanding of the history of galaxy formation and reionization, as well to our understanding of the evolution of large-scale structure in our universe~\citep{Chang10,VisLoeb10,Gong11,Pullen13,Uzgil14,BKK15b,Li15}.

Line-intensity surveys have a host of foregrounds and systematics, but a foreground of primary concern is emission from spectral lines other than the line of interest. 
Removal of this foreground presents a challenge not present in some other foregrounds, like dust or CMB emission, which are spectrally smooth across the observed frequency band. Foreground line emission, on the other hand, has sharp, structured variations in intensity, as we would expect from any spectral line. 

Here, we explore the effect of line emission contamination specifically in the context of the initial phase of the Carbon monOxide Mapping Array Pathfinder (COMAP), an experiment that is in development to map carbon monoxide (CO) line emission in CO(1-0) (rest frequency 115.26 GHz) at redshift $z\approx2.4$--3.4 (see~\autoref{tab:comapparams} for details). We expect the COMAP results to complement those of the interferometric CO Power Spectrum Survey (COPSS; see~\citealt{COPSS}).

We expect one of the brightest contaminant lines for COMAP and other CO(1-0) surveys to be a hydrogen cyanide (HCN) line, HCN(1-0), with a rest frequency of 88.63 GHz~\citep{BKK15}. CO emission from a redshift range of $z\approx2.4$--3.4, as targeted by COPSS and COMAP, mixes with HCN emission from $z\approx1.6$--2.4. In attempting to understand the effect of such contamination on CO observations, we \replaced{ask}{consider in this work}:
\begin{itemize}
\item What intensities\replaced{ do we expect}{ are expected} from HCN emission\replaced{ in a CO survey?}{?}
\item To what extent do we estimate HCN emission would \replaced{boost}{bias} the CO auto power spectrum? 
\item \replaced{What room for variation exists in our models for CO and HCN}{How do uncertainties in modelling} line emission\added{ affect predictions of HCN contamination in a CO survey}? 
\end{itemize}

This work is not the first to ask some of these questions. Indeed, it follows the works of~\cite{BKK15} and~\cite{VTL11} on line foregrounds. However, here we incorporate the methods of~\cite{Li15}\added{ to estimate the CO signal and HCN contamination}. We derive line-luminosity relations from star-formation histories in cosmological simulations and observations of many local galaxies\footnote{\cite{VTL11} calibrate many of their line-luminosity relations solely based on M82, and do not examine HCN(1-0). In addition, their focus was on CO(1-0)--CO(2-1) cross-correlation.}. \replaced{We also}{Furthermore, as in~\cite{Li15}, we} use halo catalogues from dark matter simulations \replaced{(like~\citealt{Li15} but unlike~\citealt{BKK15}) to populate our simulated survey volumes}{rather than analytically derived galaxy distributions as in~\cite{BKK15}}. \added{This approach affords greater flexibility in generating mock galaxies and sky data in preparation for COMAP, while requiring fewer simplifying assumptions about galaxy clustering.}
Our work is thus novel in the sum of its methodology and objectives\added{, and continues a larger exploration of synthetic COMAP observations that began with~\cite{Li15}}.

The paper is structured as follows: in~\autoref{sec:methods} we introduce our methods for simulating CO observations and HCN contamination in these observations, and how we can vary those methods. We then present expected contamination in observed intensity and in power spectra in~\autoref{sec:results}, with some considerations of impact on detection significance of the reduced data. After some discussion of these results and their implications for COMAP and other future CO surveys in~\autoref{sec:discuss},
we present our conclusions in~\autoref{sec:conclude}.

Where necessary, we assume base-10 logarithms, and a $\Lambda$CDM cosmology with parameters $\Omega_m=0.286$, $\Omega_\Lambda=0.714$, $\Omega_b=0.047$, $h=0.7$, $\sigma_8=0.82$, and $n_s=0.96$.  

\floattrue\begin{deluxetable}{cc}
\tabletypesize{\footnotesize}
\tablecaption{\label{tab:comapparams}
COMAP instrumental and survey parameters assumed for this work.}
\tablehead{
\colhead{Parameter} & \colhead{Value}}
\startdata
System temperature & 44 K\\
Angular resolution & $4'$\\
Frequency resolution & 40 MHz\\
Observed frequencies & 26--30 GHz\\&30--34 GHz\\
Number of feeds & 19\\
Survey area per patch & 2.5 deg$^2$\\
On-sky time per patch & 1500 hours\tabularnewline
\hline
\enddata
\tablecomments{Feeds are single-polari\replaced{s}{z}ation. The survey will cover a total range of 26--34 GHz in frequency, but with two separate downconverter systems each covering a 4 GHz band in that range. The angular resolution above is the full width at half maximum\deleted{ (FWHM) }of the Gaussian beam profile, for the receiver's central pixel.}
\end{deluxetable}

\section{Methods}
\label{sec:methods}
The methods and results of~\cite{Li15} form the basis for this study. In short, we model and apply the galaxy--halo connection to dark matter halos identified in a cosmological N-body simulation. Through a chain of empirical relations, we assign galaxy line-luminosities to individual halos, and use these luminosities to generate simulated data and power spectra. We summarise this part of our methodology in~\autoref{sec:simobs} and~\autoref{sec:dataprod}, but first in~\autoref{sec:context} we compare the scope of this work to that of previous work with similar methods and aims.

\subsection{Context in previous work}
\label{sec:context}

The work of \cite{BKK15} considers line foregrounds in the context of hypothetical CO, [CII], and Ly-$\alpha$ intensity surveys, and proposes that mitigation of these foregrounds in a CO survey is feasible by masking the brightest parts of the map. To motivate this mitigation,~\cite{BKK15} show that HCN significantly contaminates simulated CO surveys, with the HCN power spectrum sitting near or above the CO spectrum. However, the authors use an assumed galaxy power spectrum to simulate an intensity map with the expected clustering, rather than a cosmological simulation. The work in~\cite{BKK15} is also limited to HCN contamination (and its mitigation) in 2D maps, for a single observed frequency channel.\replaced{\footnote{It is not immediately obvious that high contamination in data products from 2D maps translates to equally high contamination in data products from a 3D cube, the latter of which is our concern.}}{ We must then ask whether high contamination in data products from 2D maps translates to equally high contamination in data products from a 3D cube, the latter of which is our concern. That question is a primary motivation for this work, and our results ultimately do demonstrate similar contamination in both 2D and 3D maps, but not necessarily contamination at the level claimed in~\cite{BKK15}.} 

As we will discuss in the following sections, the simulations of~\cite{Li15} incorporate complications not present in the simulations of~\cite{BKK15}. While the results of~\cite{Li15} show promising prospects for detection of the CO auto spectrum and meaningful constraints on properties of galaxy populations, our work now adds potential effects of HCN contamination as forecast by~\cite{BKK15}.
\subsection{Simulated observations}
\label{sec:simobs}

In~\autoref{sec:dmsim}, we describe the dark matter simulation used to generate the halos in our observation volume, and in~\autoref{sec:hmll} and~\autoref{sec:lumobs}, we outline the procedure used to generate simulated temperature cubes with these halos.

\subsubsection{Dark matter simulation}
\label{sec:dmsim}
We use the cosmological N-body simulation \replaced{\footnote{\texttt{c400-2048}, provided by Matthew Becker; from Becker et al., in prep..}}{{\tt c400-2048}} described in~\cite{Li15}, which provides implementation details of the simulation and subsequent halo identification. The simulation has a dark matter particle mass of $5.9\times10^8\,M_\odot\,h^{-1}$\added{, and we include dark matter halos more massive than $M_\text{vir} = 10^{10}\,M_\odot$ in our analysis. We will address this mass cutoff further in~\autoref{sec:hmll} and~\autoref{sec:hmll_mvar}}. 

To simulate galaxies in our field of observation, we use dark matter halos identified in ``lightcone'' volumes, enclosing all halos within a given sky area and redshift range, with each lightcone based on arbitrary choices of observer origin and direction within the cosmological simulation. We use 100 $z=2.3$--2.9 and 100 $z=1.5$--2.0 lightcones populated with $\sim10^6$--$10^7$ halos over a flat-sky area of $100'\times100'$. Random pairings of these lightcones form the basis for our simulated observations.

\subsubsection{Halo mass--line-luminosity relation}
\label{sec:hmll}
We assign each halo a luminosity in each line based on its sky position, mass, and cosmological redshift (excluding peculiar velocities). We first consider a fiducial model (the `\replaced{turnabout}{turnaround} model') that builds on previous work in~\cite{Li15}, and then consider variations on this model in the following section.

\paragraph{Fiducial model}
The ladder of relations used in~\cite{Li15} to convert halo masses to CO line-luminosity is as follows.
\begin{itemize}
\item Assume only halos above a cutoff mass of $10^{10}\ M_\odot$ can support emission in any line, and assign zero luminosity to halos below this mass.
\item Convert halo masses to star-formation rates (SFR) for each halo via interpolation of data from~\cite{Behroozi13a} and~\cite{Behroozi13b}. The main focus of these papers is to constrain the stellar mass--halo mass relation and derived quantities such as SFR by comparing simulation data with observational constraints, and the resulting data include average SFR in a halo given its mass and redshift. This relation shows a peak in SFR around $10^{11.7}\ M_\odot$ on average, beyond which star-formation efficiency appears to fall off. This \replaced{turnabout}{turnaround} in the relation is consistent with previous analyses, as~\cite{Behroozi13a} note.
\item Approximate halo-to-halo scatter in SFR by adding 0.3 dex log-normal scatter to the SFR obtained above, preserving the linear mean.
\item Convert SFR into infrared (IR) luminosity, using a known tight correlation:
\begin{equation}\frac{{\rm SFR}}{M_\odot\,\text{yr}^{-1}} = \delta_\text{MF}\times10^{-10}\left(\frac{L_\text{IR}}{L_\odot}\right).\end{equation}
As in~\cite{Behroozi13a} and~\cite{Li15}, we take $\delta_\text{MF}=1.0$, corresponding to a Chabrier initial mass function (IMF). We note the uncertainty of this relation in low-mass halos, where dust obscuration may not completely cover star-formation activity~\citep{WuDore17}.
\item Convert between IR luminosity and observed CO luminosity through power-law fits to observed data, commonly given in the literature:
\begin{equation}\log{\left(\frac{L_\text{IR}}{L_\odot}\right)}=\alpha\log{\left(\frac{L'_\text{CO}}{\text{K km s}^{-1}\text{ pc}^2}\right)}+\beta,\end{equation}
where for our fiducial model, we take $\alpha=1.37$ and $\beta=-1.74$ from a fit to high-redshift galaxy data ($z\gtrsim1$) given in~\cite{CW13}, following~\cite{Li15}.
\item Add 0.3 dex log-normal scatter in CO luminosity, again preserving the linear mean.
\end{itemize}

In general, to convert halo masses to \emph{any} line-luminosity that correlates reasonably tightly with IR luminosity, we simply need a relation of the form
\begin{equation}\log{\left(\frac{L_\text{IR}}{L_\odot}\right)}=\alpha\log{\left(\frac{L'_\text{line}}{\text{K km s}^{-1}\text{ pc}^2}\right)}+\beta.\end{equation}
\cite{GS04}, based on a mixed sample of normal galaxies and luminous and ultra-luminous infrared galaxies (LIRGs and ULIRGs, $z<0.06$), obtain $\alpha=1.00\pm0.05$ and $\beta=2.9$ for HCN(1-0). Since $\alpha=1$, this relation between IR and HCN luminosity is linear.

$L'_\text{line}$ is the observed luminosity (or velocity- and area-integrated brightness temperature) of the halo, which we convert into an intrinsic luminosity for each halo, as in~\cite{Li15}:
\begin{equation}\frac{L_\text{line}}{L_\odot}=4.9\times10^{-5}\left(\frac{\nu_\text{line,rest}}{115.27\text{ GHz}}\right)^3\frac{L'_\text{line}}{\text{K km s}^{-1}\text{ pc}^2}.\end{equation}
Note that while this relation is taken from a CO model, it is actually a re-statement of
\begin{equation}L_\text{line}=\frac{8\pi k_B}{c^3}\nu_\text{line,rest}^3L_\text{line}',\end{equation}
which arises from general definitions of the two luminosity quantities in the Rayleigh--Jeans approximation, which we still apply for HCN line emission.\footnote{Emission in CO(1-0) and HCN(1-0) is not technically in the Rayleigh--Jeans regime, but the historical convention in radio astronomy is to report brightness temperature in this approximation, where `temperature' scales linearly with surface brightness~\citep{CW13}. We continue to use this convention, which allows for easier comparison not only with previous literature, but with instrument sensitivity.}

In simulating halo-to-halo scatter in the line-luminosity itself, we add 0.3 dex log-normal scatter to both CO and HCN line-luminosities in the IR--line-luminosity relation. In both lines, the amount of scatter is consistent with the amount of scatter present in the local observations forming the basis for our scaling relations. Whether this scatter accurately reflects fluctuations in star-formation activity in individual galaxies is an outstanding question, but we will at least attempt to address how our results depend on the amount of scatter in the following sections.

\subsubsection{Model variations}
\label{sec:hmll_mvar}
We introduce a few variations on the fiducial model, in order to address potential concerns:
\begin{itemize}
\item Various papers since~\cite{GS04} have carried out re-analyses of the same data with small additional samples (mostly at $z\ll1$, but also a handful of detections at $z\gtrsim1$). The re-analyses in~\cite{Carilli05} and~\cite{GB12} show the IR--HCN correlation to be at least marginally non-linear, respectively with best-fit parameters of $(\alpha,\beta)=(1.09\pm0.02,2.0)$ and $(\alpha,\beta)=(1.23\pm0.05,1.07\mp0.40)$. Since we assume the linear relation from~\cite{GS04}, we should understand how our estimates of contamination vary with the mean IR--HCN relation.
\item The scatter we introduce in SFR and in HCN luminosity is shorthand for variations in star-formation activity and gas dynamics in individual halos. The fiducial values we assume are based on local observations, and cosmic trends in such dynamical activity may result in greater halo-to-halo scatter at some redshifts than at others. We thus explore how our estimates of contamination vary with increased scatter in HCN luminosity.
\item In their work,~\cite{BKK15} use a single power-law relation to assign luminosities to halos, very unlike our fiducial model which introduces a \replaced{turnabout}{downturn} at high masses. We must explore how this simplified form of the halo mass--SFR relation affects our estimates of HCN contamination.
\end{itemize}

\paragraph{The mean $L_{\rm IR}$--$L_{\rm HCN}'$ relation}
For this paper, we simply generate additional halo luminosities for each lightcone, based on the alternate $\alpha$ and $\beta$ from the two re-analyses mentioned above. These alternate values do not vary by more than a factor of order unity from the values given in~\cite{GS04}. We may not expect this level of variation to make much difference in the HCN power spectrum, especially relative to CO.
We will quantitatively show in~\autoref{sec:res_halotemp}\added{,~\autoref{sec:res_pspec},} and~\autoref{sec:SNR} that indeed, such variation makes little difference in the HCN temperatures and CO spectrum contamination, especially compared to the other model variations listed here.

\cite{GS04} also suggest a strong CO--HCN correlation in luminosity, based on a fit of HCN luminosities to both CO and far-IR luminosities. Again, given that the above variations in determining HCN luminosity make little difference, we omit consideration of CO-dependence for simplicity.

\paragraph{Halo-to-halo scatter in line-luminosity}

We have no reason to suspect significant cosmic evolution of scatter in luminosities with redshift based on observational data. \cite{Sun16} also study line emission contamination in line-intensity mapping (in the context of the Tomographic Ionised-carbon Mapping Experiment, or TIME, designed to map [CII] emission at $z\sim6.5$), and mitigation through masking brighter foreground galaxies. As part of this work, that paper characterises the halo mass--$L_{\rm IR}$ relation, and sees no obvious evolution of scatter in $L_{\rm IR}$ with redshift up to $z\sim1.5$.

However, we should still explore the possibility that scatter in the $L_{\rm IR}$--$L'_{\rm line}$ relation \emph{does} vary over redshift or across different gas species. We vary the HCN luminosity scatter between 0.0 dex, 0.3 dex, 0.5 dex, and 1.0 dex. Of these, 0.3 dex is the fiducial value, and the value used with the \replaced{turnabout}{turnaround} model unless otherwise specified. Note that we keep the SFR scatter at 0.3 dex in all simulations.

\paragraph{The form of the halo mass--SFR relation}
The power-law model of~\cite{BKK15} follows~\cite{Pullen13}, which in part investigates cross-correlation prospects between CO line emission and samples of quasi-stellar objects (QSO). The model is as follows:
\begin{itemize}
\item Select a fraction of halos to emit in each line, based on the assumption of a duty cycle for emission. The model assumes that the duty cycle of star-formation and thus line emission is given by \begin{equation}f_\text{duty}=t_s/t_\text{age}(z),\end{equation} where $t_s\sim10^8$ years is the approximate time-scale of star-formation, and $t_\text{age}(z)$ the age of the universe in our redshift range. While the duty cycle thus evolves with redshift, we take a constant $f_\text{duty}$ for each lightcone, using the average age of the universe over the applicable redshift range.
\item For those halos that do emit, the halo mass--line-luminosity relation per halo is a simple power law with empirically constrained parameters:
\begin{equation}\frac{L_\text{line}}{L_\odot}=A\left(\frac{M}{M_\odot}\right)^b,\end{equation}
where for CO,~\cite{BKK15} use $A=2\times10^{-6}$ and $b=1$ following~\cite{Pullen13} (originally from~\citealt{Wang10}), and for HCN use $A=1.7\times10^{-15}$ and $b=1.67$ as derived from~\cite{GS04}.
\end{itemize}
Given $t_s=100$ Myr as assumed in past literature, $f_\text{duty}\approx2.5\%$ for the HCN lightcones and $\approx4\%$ for the CO lightcones. However, observations at $z\sim1$--3 suggest a near-unity duty cycle~\citep{Daddi07,Noeske07,COPSS}. Therefore, we use two power-law models with different uses of $f_\text{duty}$. One model follows~\cite{BKK15} closely, selecting $f_\text{duty}$ of halos above the cutoff mass to emit with luminosity $A(M/M_\odot)^bL_\odot$. In the other model, we assume a duty cycle of unity and absorb $f_\text{duty}$ into the luminosity calculation, so that all halos above the cutoff mass emit with luminosity $f_\text{duty}A(M/M_\odot)^bL_\odot$. 

\begin{figure}[t]
\includegraphics[width=\linewidth]{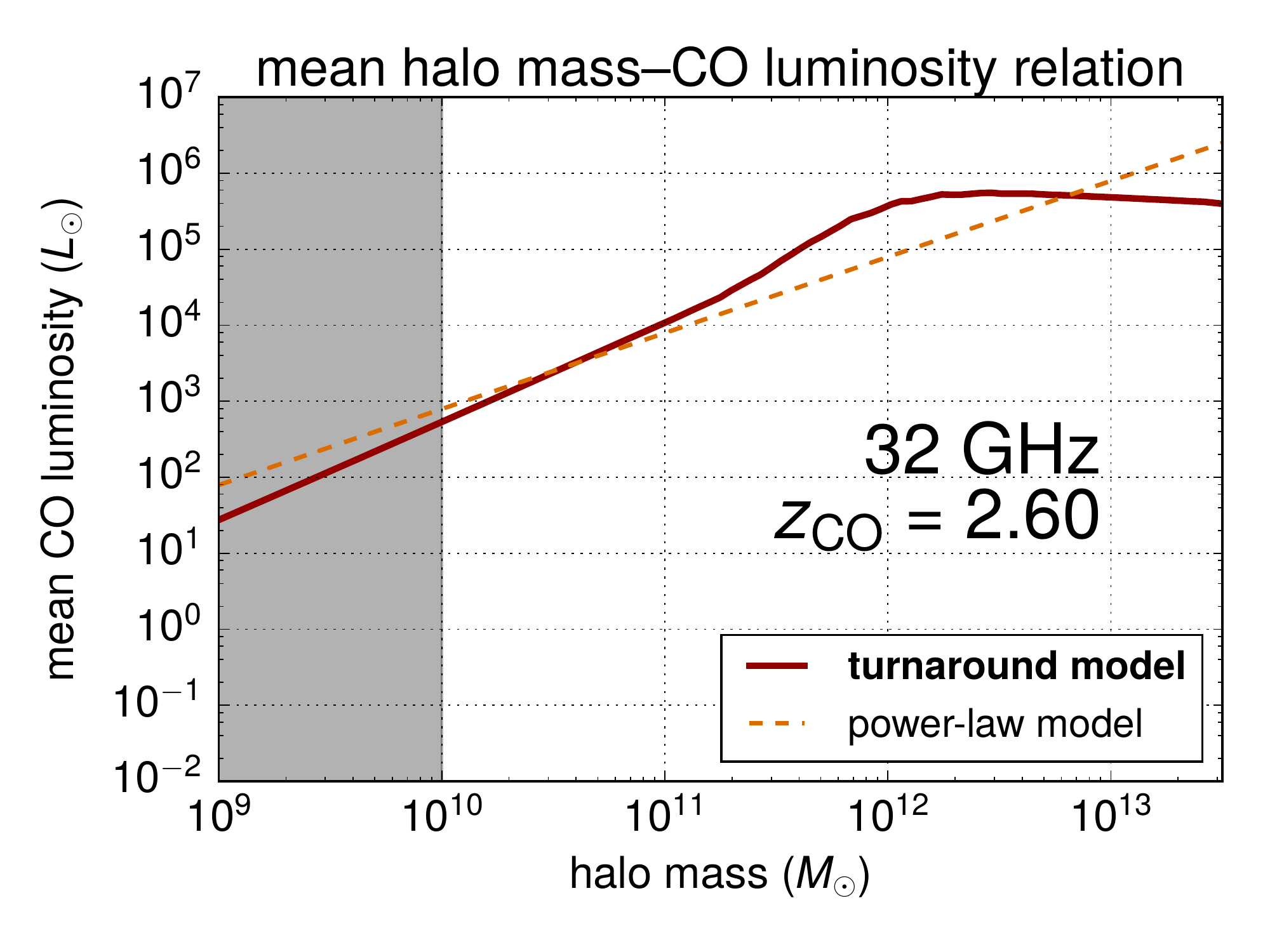}

\includegraphics[width=\linewidth]{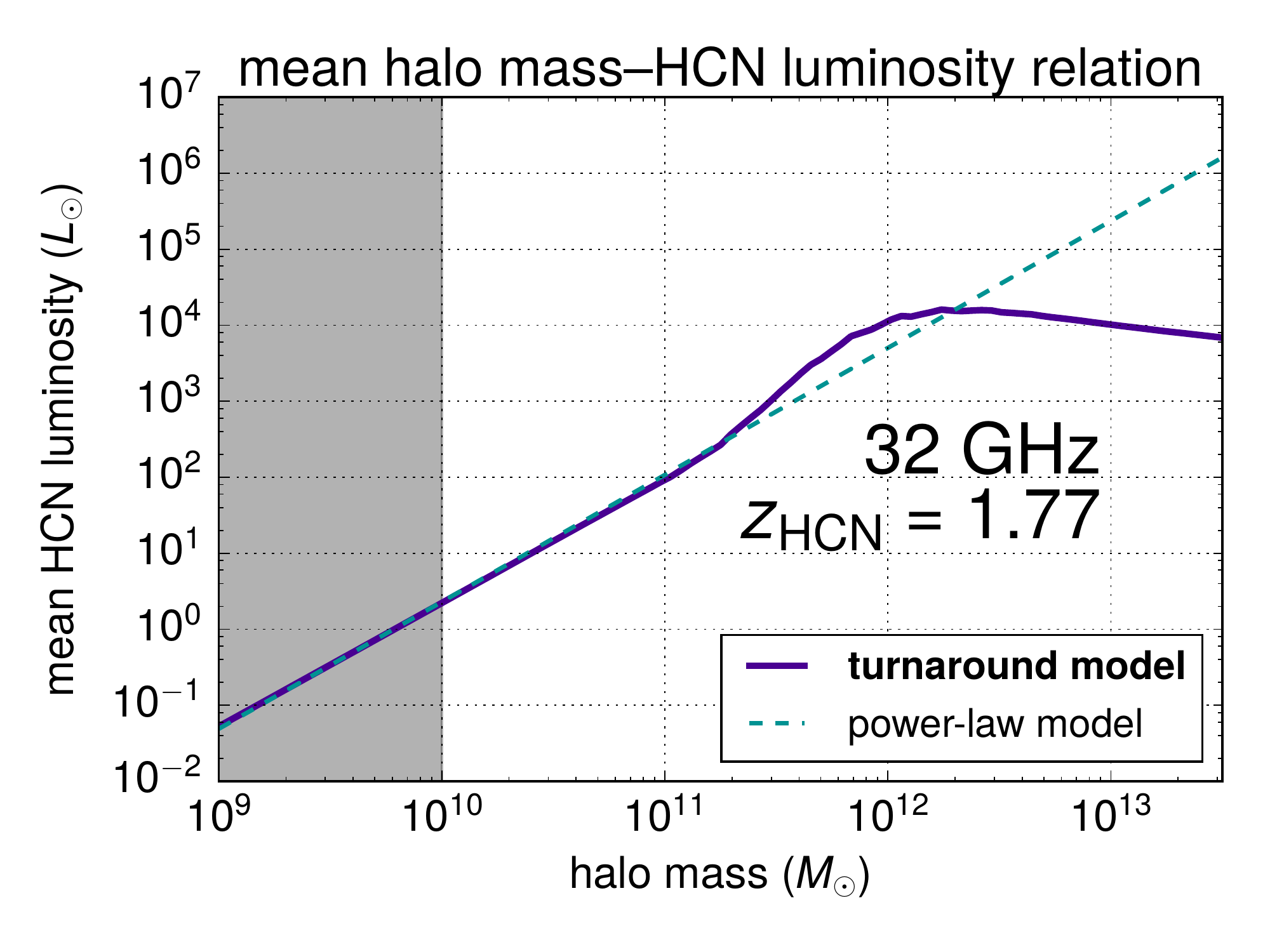}
\caption{Mean relations for halo mass--CO luminosity at $z=2.60$ ({\bf upper panel}) and halo mass--HCN luminosity at $z=1.77$ (\textbf{lower panel}), used in this work and in~\cite{BKK15}. 
The redshifts correspond to an observed frequency of 32 GHz for each line. Simulations assign luminosities with halo-to-halo scatter not depicted in this figure. For the power-law model, the relations plotted include the factor of $f_\text{duty}$ for CO and HCN. The grey area in each panel indicates halo masses excluded from emitting at all in this work and in~\cite{Li15}, but not in~\cite{BKK15}.}
\label{fig:halol}
\end{figure}

\capstartfalse
\floattrue\begin{deluxetable}{cl}
\tabletypesize{\footnotesize}
\tablewidth{0.9\linewidth}
\tablecaption{\label{tab:modelsum}
Summary of HCN models explored in this paper.}
\tablehead{
\colhead{Model} & \colhead{Description}}
\startdata
\textbf{\replaced{turnabout}{turnaround}}&based on~\cite{Li15}
\\&\textbullet\enskip SFR from \cite{Behroozi13a} relation
\\&\qquad (peaks at halo masses of $\sim10^{12}$ $M_\odot$)
\\&\textbullet\enskip log-normal scatter in SFR of 0.3 dex
\\&\textbullet\enskip SFR scaled to $L_{\rm IR}$ assuming IMF
\\&\textbullet\enskip $L_{\rm IR}$--$L_{\rm HCN}$ scaling from~\cite{GS04}
\\&\textbullet\enskip scatter in $L_{\rm HCN}$ of 0.0, \textbf{0.3}, 0.5, or 1.0 dex
\\
\hline
power-law&based on~\cite{BKK15}
\\&\textbullet\enskip $L_{\rm HCN}/L_\odot=A(M/M_\odot)^b$\\&\textbullet\enskip only $f_{\rm duty}$ of halos emit in HCN\\
&\emph{alternatively:}\\
&\textbullet\enskip $L_{\rm HCN}/L_\odot=f_{\rm duty}A(M/M_\odot)^b$\\&\textbullet\enskip all halos emit in HCN line\\
\hline
\enddata
\tablecomments{The fiducial model parameters are indicated in bold. In all cases, there is a cutoff mass below which halos do not host HCN emitters, and that mass is $10^{10}$ $M_\odot$ in this paper for all models.}
\end{deluxetable}
\capstarttrue

Our fiducial model HCN relation is based on the same IR--HCN luminosity relation that~\cite{BKK15} use. The CO relation used for the fiducial model does differ from that used for the power-law models and in~\cite{BKK15}, 
but we use those values for our power-law models to compare our results more easily with~\cite{BKK15}. Broadly speaking, $\alpha$ (or $(5/3)b^{-1}$) for CO is still significantly higher than $\alpha$ for HCN, resulting in lower-mass halos contributing more in CO emission than in HCN emission.

\autoref{fig:halol} shows the mean halo mass--line-luminosity relations at the redshifts for CO and HCN corresponding to the midpoint of the COMAP 30--34 GHz band. 
Due to the \replaced{turnabout}{turnaround} in the halo mass--SFR relation from~\cite{Behroozi13a}, the average line-luminosity for halos of mass $\gtrsim10^{12}\ M_\odot$ is substantially lower in our fiducial model than in either power-law model, especially for HCN. We \replaced{suggest}{expect} that the lower luminosities are more realistic, based on \replaced{primitive}{(a) the clear constraints on the shape of the stellar mass--halo mass relationship in~\cite{Behroozi13a} and (b)} comparisons with existing \added{HCN} observations in~\autoref{sec:obsHCN}.

Simulations with our power-law models still diverge in important details from~\cite{BKK15}. We use simulated dark matter halos in 3D rather than generating 2D maps to fit expected power spectra. The particle mass of our dark matter simulation ($\sim10^9\ M_\odot$) limits the mass-completeness of our halo catalogues, so we continue to use a cutoff mass for emission of $10^{10}\ M_\odot$, rather than the cutoff mass in~\cite{BKK15} of $10^9\ M_\odot$. So we treat these power-law models as a variation of the form of the halo mass--SFR relation, not an attempt to replicate the results of~\cite{BKK15}.

A lower cutoff mass cuts out a potential population of faint emitters, impacting the signal and contaminant forecasts (and we will return to this point in~\autoref{sec:res_halotemp} and~\autoref{sec:bkk15_comp}). That said, the impact would not be at the scale of orders of magnitude. Take the average map temperature as a zeroth-order metric of signal intensity. \cite{Li15} (in Appendix A.1 of that paper) make an analytic calculation of $\avg{T_{\rm CO}}$ as a function of the minimum CO-emitting halo mass, based on the fiducial halo mass--CO luminosity scaling relations and an assumed halo mass function (not subject to completeness concerns). Through this calculation, they show that moving the cutoff mass down to $10^9\ M_\odot$ would not have significantly impacted their mean CO brightness temperature $\avg{T_{\rm CO}}$ for the fiducial model. Since HCN luminosity falls even faster with decreasing halo mass (see~\autoref{fig:halol}), we expect changes in cutoff mass to affect HCN even less.

For easy reference, we recap the fiducial and power-law models in~\autoref{tab:modelsum}.

\subsubsection{Halo luminosities to observations}
\label{sec:lumobs}

\autoref{tab:comapparams} gives the instrumental and survey parameters anticipated for the initial phase of COMAP. We use these parameters to generate a data cube from the halo luminosities assigned within the lightcone.

Each volume element (voxel) in the data cube has angular widths $\delta_x$ and $\delta_y$ and frequency resolution $\delta_\nu$. Here, $\delta_x=\delta_y=0.6$', while $\delta_\nu=40$ MHz.\footnote{This corresponds to a comoving voxel of roughly $1.1\times1.1\times4.6$ cMpc$^3$ at redshift 2.60, or $0.86\times0.86\times5.6$ cMpc$^3$ at redshift 1.77.} (Following changes in COMAP design during the preparation of this work, we now expect the actual experiment to take data with $\delta_\nu\approx10$ MHz.) Since the sky grid is an order of magnitude finer than the stated angular resolution of the actual survey, the simulated maps retain modes on spatial scales smaller than would be observable in reality. This should not affect any conclusions we draw about the level of contamination that HCN emission would contribute to the CO signal.

\begin{figure*}[t]
\centering\includegraphics[width=0.942\linewidth]{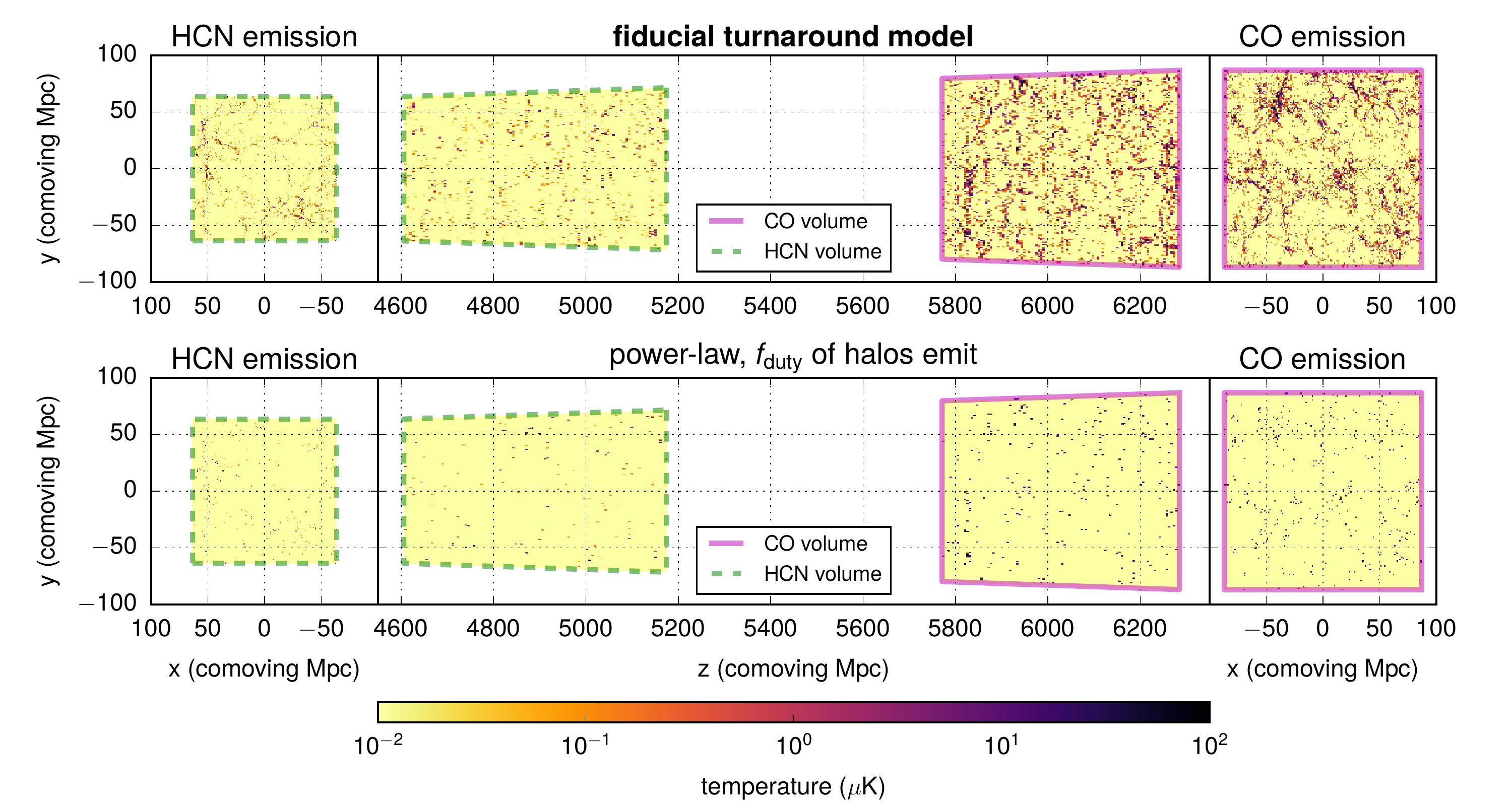}
\caption{Sample slices from temperature cubes generated with the fiducial model ({\bf upper panels}) and a power-law model ({\bf lower panels}), projected into comoving space. The underlying halos are the same. {\bf Left panels:} a slice of the HCN temperature cube, in transverse (on-sky) directions. {\bf Middle panels:} slices of the HCN and CO temperature cubes, with one axis along the line of sight. Note the difference in comoving volume and location. {\bf Right panels:} a slice of the CO temperature cube, again in transverse directions.}
\label{fig:slice_viz}
\end{figure*}

We follow~\cite{Li15} again in generating a temperature cube:
\begin{itemize}
\item Bin the halo luminosities into resolution elements in frequency and angular position, resulting in a certain luminosity associated with each voxel.
\item Convert these luminosities into surface brightness (apparent spectral intensity, in units of luminosity per unit area, per unit frequency, per unit solid angle):
\begin{equation}I_{\nu,\text{obs}} = \frac{L_\text{line}}{4\pi D_L^2}\frac{1}{\delta_x\delta_y\delta_\nu}.\end{equation}
\item Convert to the expected brightness temperature contribution from each voxel. The Rayleigh--Jeans brightness temperature for a given surface brightness is
\begin{equation}T=\frac{c^2I_{\nu,\text{obs}}}{2k_B\nu_\text{obs}^2},\end{equation}
from which we obtain our temperature for each voxel in the data cube.
\end{itemize}

At the time of writing, we do not have a clear idea of what each COMAP sky patch will look like, so we simulate an observation over a square $95'\times 95'$ patch in a 30--34 GHz frequency band (corresponding to our choice of lightcones covering $z=2.3$--$2.9$ and $z=1.5$--$2.0$). Note that during the preparation of this work, COMAP expanded its instrument bandwidth to cover 26--30 GHz as well. We do not expect such differences from the actual survey, or the omission of various systematics and cosmological effects, to radically affect our conclusions.





For a given pair of lightcones, we simulate CO and HCN temperature data in the same observed sky and frequency bins, and then add them together to simulate a CO observation with HCN contamination. This addresses any concerns discussed in~\cite{BKK15} that arise from projecting power spectra from a lower-redshift region into a higher-redshift region.~\autoref{fig:slice_viz} depicts the difference in the comoving volumes observed in CO(1-0) versus in HCN(1-0) by the same survey.


\subsection{Power spectra from observations}
\label{sec:dataprod}

We work with the spherically averaged, comoving 3D power spectrum $P(k)$, and the average of all 2D sky angular power spectra $C_\ell$ from each frequency channel in the data cube. We calculate $P(k)$ by calculating the full 3D power spectrum $P(\mathbf{k})$ of the temperature cube in comoving space, binning it in radial `shells' of $k=|\mathbf{k}|$, and averaging the full spectrum in each bin. We obtain $C_\ell$ in each frequency channel through similar binning of the 2D power spectrum in $\mathbf{k}$-space. As~\autoref{fig:slice_viz} shows, the comoving volume is not perfectly rectangular, but we approximate it as such for $P(k)$ calculations here.

As in~\cite{Li15} and~\cite{BKK15}, we present the spherical 3D power spectra as $\Delta^2(k)=k^3P(k)/(2\pi^2)$, and the averaged 2D power spectra in the conventional $C_\ell$ form, using a flat-sky approximation as in~\cite{Chiang12} and~\cite{BKK15}.\added{ While the 3D analogue to $C_\ell$ is just $P(k)$ rather than the rescaled $\Delta^2(k)$, and the 2D analogue to $\Delta^2(k)$ is the suitably rescaled $\ell(\ell+1)C_\ell/(2\pi)$ rather than just $C_\ell$, we use these rather disparate presentations to follow the conventions used in previous literature.}

\capstartfalse\floattable
\begin{deluxetable}{cccc}
\tabletypesize{\footnotesize}
\tablewidth{0.9\linewidth}
\tablecaption{\label{tab:meantemps}
Median and 95\% sample interval of mean brightness temperature in each line over the survey volume, in simulated observations of all 496 lightcone pairs.}
\tablehead{
\colhead{Model description} & \colhead{$\sigma_{L_{\rm HCN}}$ (dex)} & \colhead{$\avg{T_{\rm CO}}$ ($\mu$K)} & \colhead{$\avg{T_{\rm HCN}}$ ($\mu$K)}}
\startdata
\multirow{4}{*}{\textbf{\replaced{turnabout}{turnaround} model}: non-power-law $M_{\rm halo}$--$L_{\rm line}$ relations, with log-normal scatter}&0.0&$0.904^{+0.125}_{-0.100}$&$0.062^{+0.009}_{-0.007}$\\
&{\bf 0.3}&$0.906^{+0.124}_{-0.102}$&$0.062^{+0.009}_{-0.007}$\\
&0.5&$0.906^{+0.120}_{-0.101}$&$0.062^{+0.009}_{-0.007}$\\
&1.0&$0.905^{+0.121}_{-0.102}$&$0.060^{+0.015}_{-0.009}$\\
\hline
power-law model, $f_{\rm duty}$ of halos with $M_{\rm h}>10^{10}M_\odot$ emit at $L_{\rm line}=A(M/M_\odot)^bL_\odot$&n/a&$0.471^{+0.064}_{-0.053}$&$0.137^{+0.121}_{-0.050}$\\
power-law model, all halos with $M_{\rm h}>10^{10}M_\odot$ emit at $L_{\rm line}=f_{\rm duty}A(M/M_\odot)^bL_\odot$&n/a&$0.470^{+0.062}_{-0.052}$&$0.146^{+0.044}_{-0.035}$\tabularnewline
\hline
\enddata
\tablecomments{The fiducial model parameters are indicated in bold. See \autoref{sec:hmll} for details of each model.  $\sigma_{L_\text{HCN}}$ does not include log-normal scatter in SFR, and is not an applicable parameter for the power-law models.}
\end{deluxetable}
\capstarttrue

\section{Results}
\label{sec:results}
\subsection{Simulated temperature cubes}
\label{sec:res_halotemp}

\begin{figure*}[t]
{\centering
\includegraphics[width=0.48\linewidth]{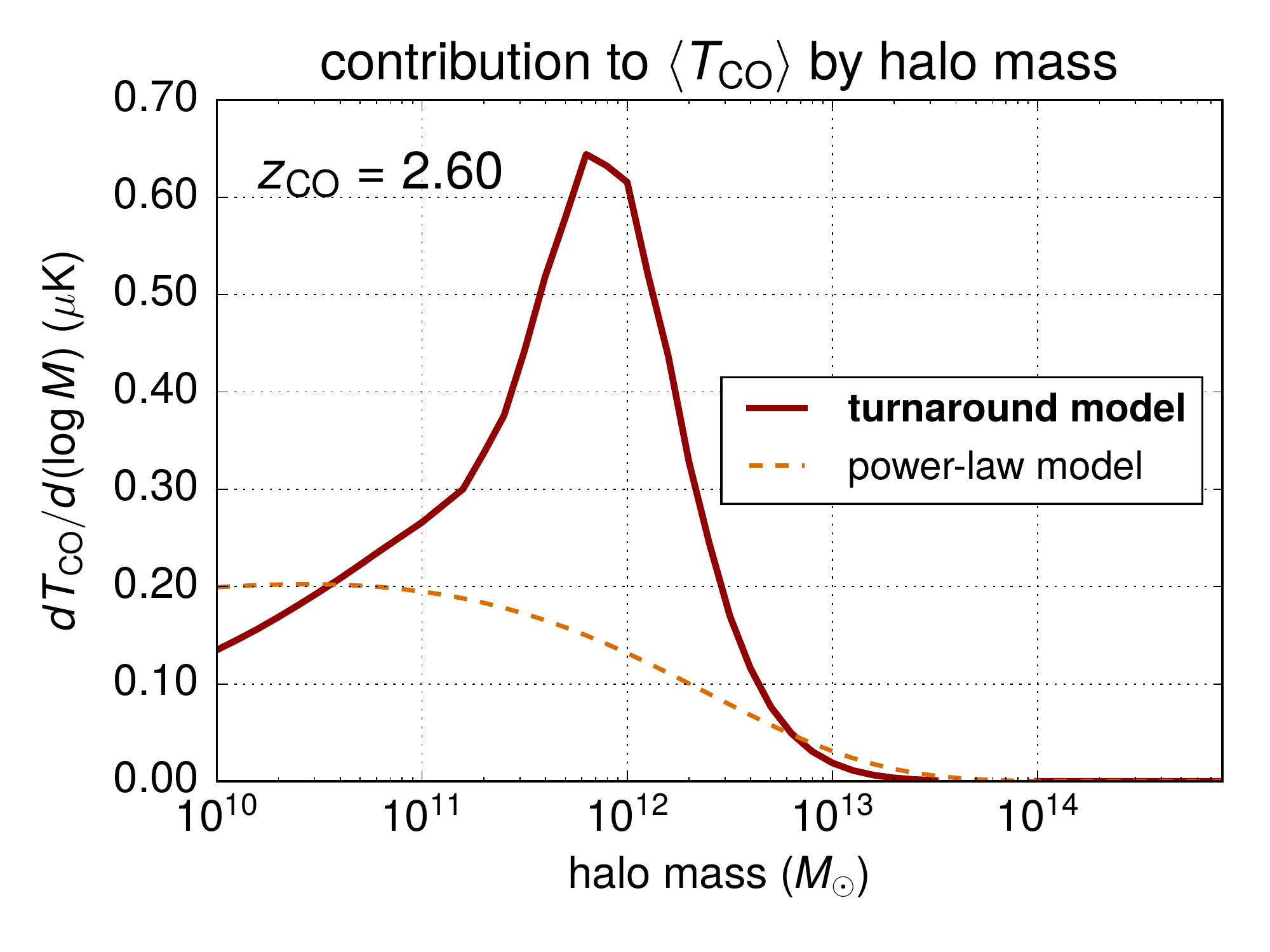}\added{\quad\includegraphics[width=0.48\linewidth]{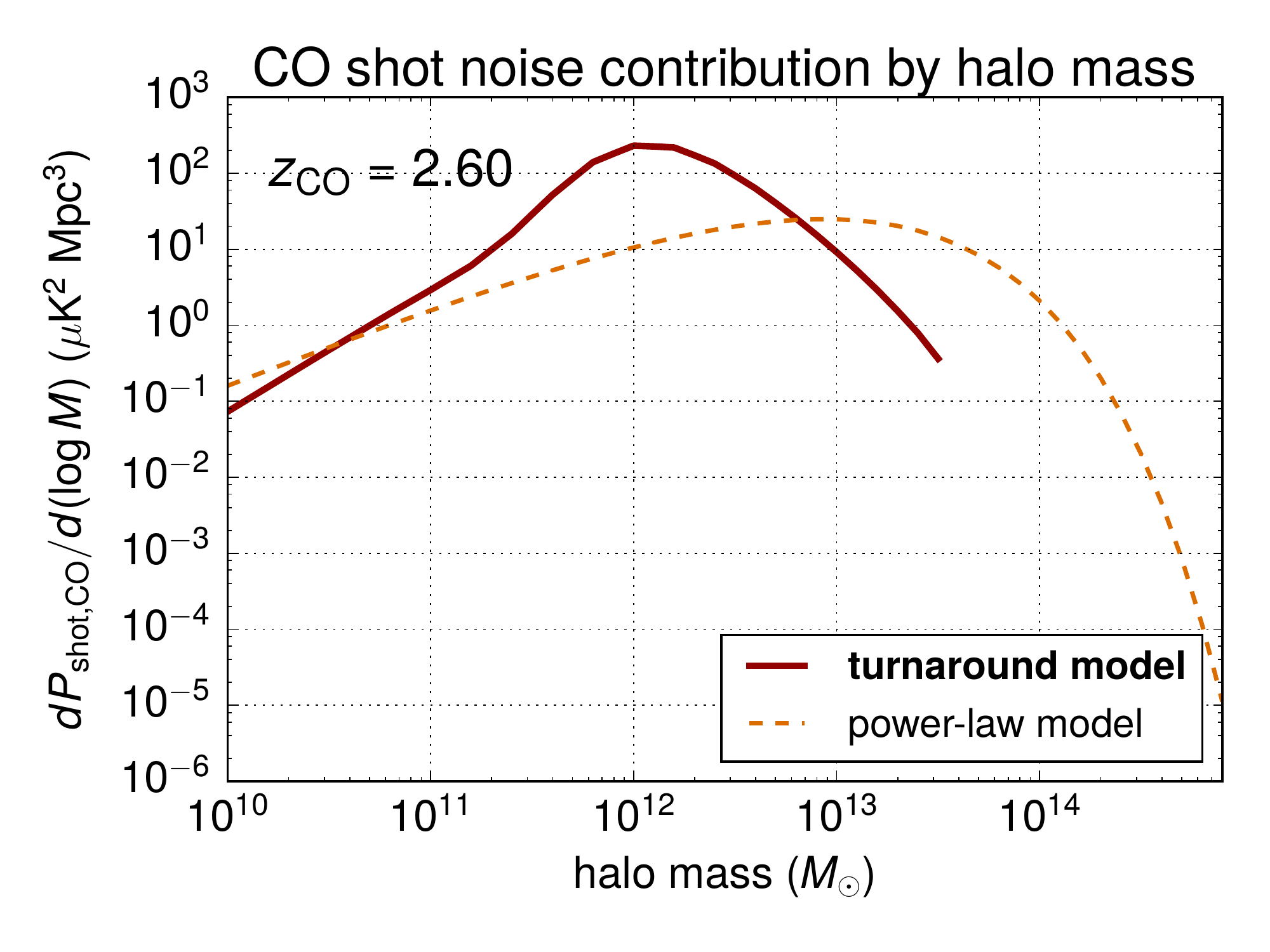}}

\includegraphics[width=0.48\linewidth]{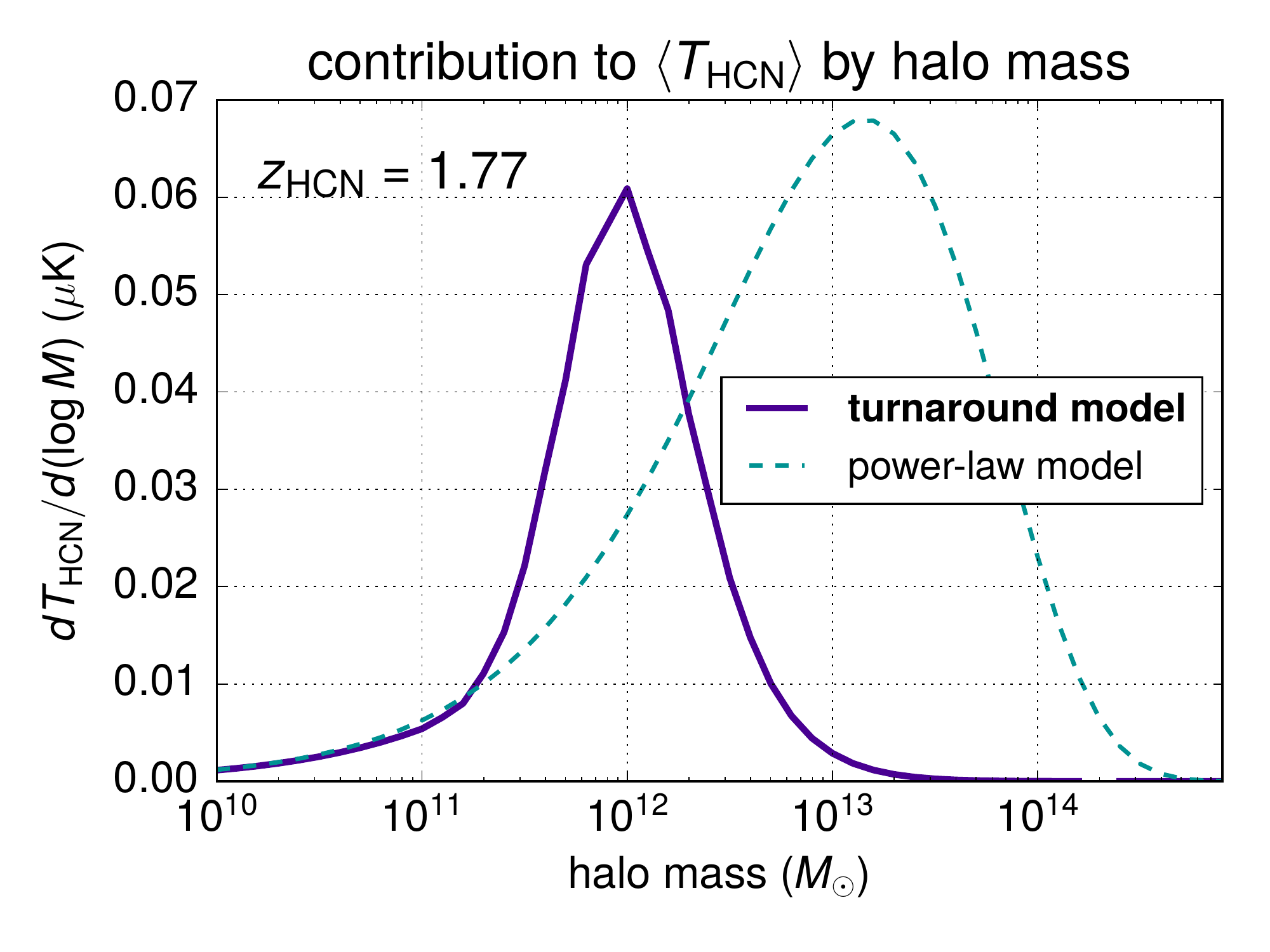}\added{\quad\includegraphics[width=0.48\linewidth]{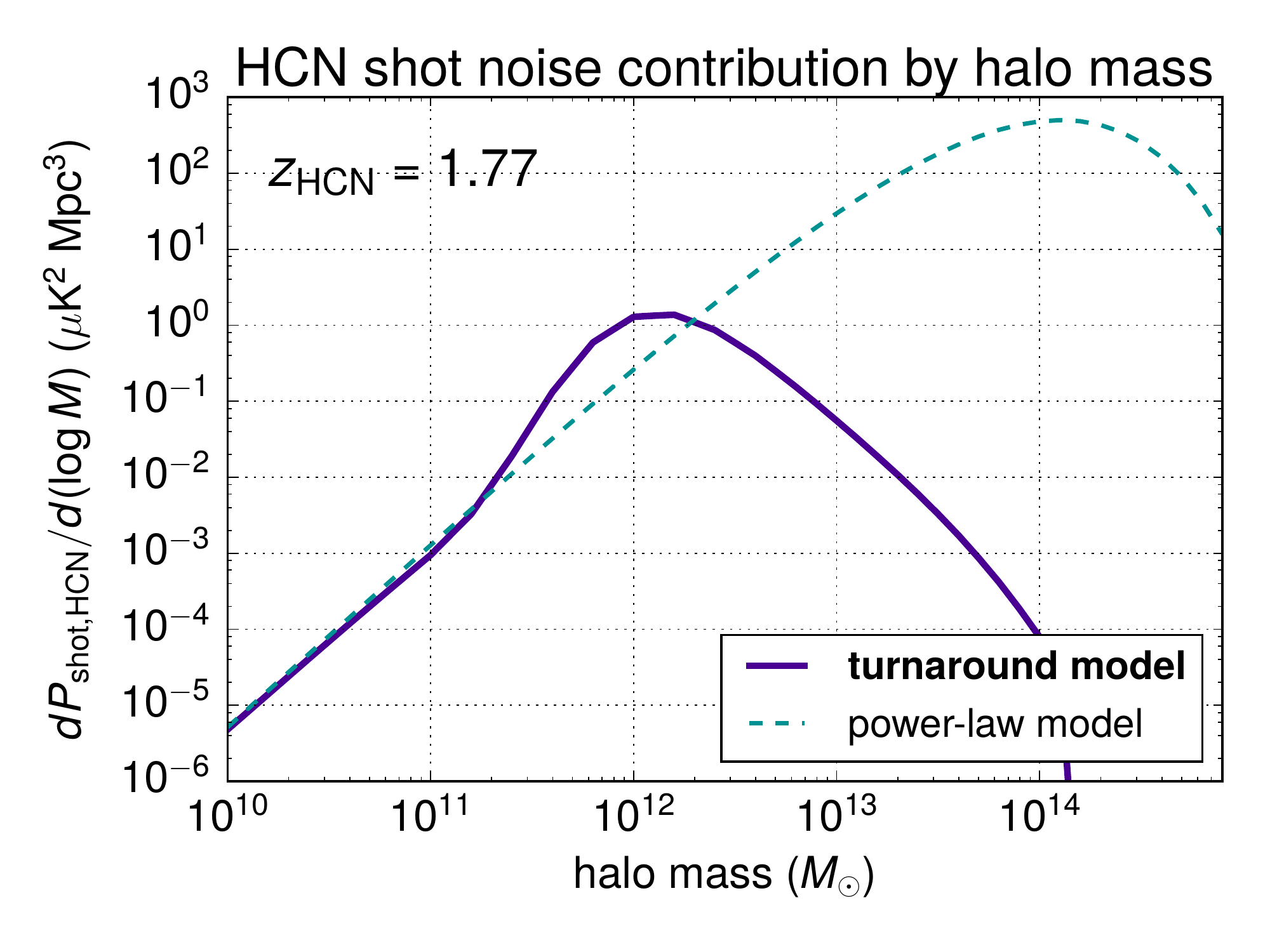}}}
\caption{Expected contributions by halo mass to brightness temperature\added{ (\textbf{left panels}) and the shot noise component of the power spectrum (\textbf{right panels}),} of CO emission at $z=2.60$ (\replaced{\bf upper panel}{\textbf{upper panels}}) and HCN emission at $z=1.77$ (\textbf{lower panel\added{s}}), for the fiducial \replaced{turnabout}{turnaround} model and the power-law model. We quantify the contribution from halo masses $M_\text{vir}\in[M,M+dM]$ \added{to average temperature} as \replaced{$dT/d(\log{M})$, converted from $L_\text{line}(M)\,dn/d(\log{M})$ calculated using}{$dT/d(\log{M})\propto L_\text{line}(M)\,dn/d(\log{M})$, and the contribution to the shot noise component as $dP_\text{shot}/d(\log{M})\propto L_\text{line}^2(M)\,dn/d(\log{M})$. Note that the y-axis limits differ between the upper left and lower left panels, but not the upper right and lower right panels. All calculations use} the line-luminosity models\text{ in this work} and the halo mass function fit from~\cite{Behroozi13b}. As in~\autoref{fig:halol}, the redshifts correspond to an observed frequency of 32 GHz for each line, at the midpoint of the simulated observing band.} 
\label{fig:dtdm}
\end{figure*}

For 496 different pairs of CO and HCN lightcones, we generate temperature cubes for CO emission and HCN emission, which we then add together in observed voxels to obtain a simulated map for CO emission plus HCN contamination. Approximately $1.5\times10^6$ halos in the CO lightcones and approximately $5.4\times10^6$ halos in the HCN lightcones typically fall above the cutoff mass of $10^{10}$ $M_\odot$ and within a $95'\times95'$ patch. Since we used only 100 lightcones populated with CO emitters and 100 lightcones populated with HCN emitters, there is some redundancy in the lightcones used. But since temperature maps were generated anew for each new pairing, even two maps from the same lightcone are subject to some differences due to re-application of halo-to-halo scatter in SFR and luminosity. On top of the halo-to-halo scatter in each lightcone, sample variance in large-scale structure between lightcones results in lightcone-to-lightcone scatter in temperature and power spectra.

\autoref{fig:slice_viz} shows slices of a sample pair of temperature data cubes for a given pair of lightcones. Note that these slices are generated with the fiducial model (the \replaced{turnabout}{turnaround} model with 0.3 dex log-scatter in luminosity). The power-law model maps appear much sparser than maps from any of the other models, especially when only $f_{\rm duty}$ of sufficiently massive halos host emitters.

\autoref{tab:meantemps} lists the mean line temperatures over each simulated CO map and HCN map. Variations in $\sigma_{L_{\rm HCN}}$ do not appear to significantly affect $\avg{T_{\rm HCN}}$, and any differences present could be ascribed to map-to-map scatter. The power-law models yield lower mean CO temperatures by approximately a factor of 2 in comparison to the fiducial model results, but also higher HCN mean temperatures by approximately a factor of 2. We might expect the slight change in the choice of empirical IR--line-luminosity relation to change the CO temperature, but there is no such change for HCN.

We can compare these cube temperatures to analytic calculations based on the line-luminosity per volume $dL_\text{line}/dV = \int L_\text{line}(M)\,(dn/dM)\,dM$, where $dn/dM$ is the halo mass function fit in~\cite{Behroozi13b} adapted to the appropriate redshifts (at the midpoint of the simulated observing band) and cosmology. The $L_\text{line}(M)$ function used here is an average across all halos of mass $M$, and thus absorbs any factors of $f_\text{duty}$. Those calculations predict $\avg{T_\text{CO}}=0.888$ $\mu$K and $\avg{T_\text{HCN}}=0.058$ $\mu$K for the fiducial model, versus $\avg{T_\text{CO}}=0.459$ $\mu$K and $\avg{T_\text{HCN}}=0.125$ $\mu$K for the power-law model. All of these temperatures are consistent with the range of empirically obtained brightness temperatures in~\autoref{tab:meantemps}. This analytic calculation also allows us to estimate the contribution to the brightness temperature from different bins of halo masses. We quantify this in~\autoref{fig:dtdm} using a slight variation on the integrand, $L_\text{line}(M)\,dn/d(\log{M})$.\added{

As a prelude to the discussion of power spectra in the next section, we also provide an analytic illustration of how different bins of halo masses contribute to the shot noise component of the power spectrum. To illustrate the contribution to mean line temperature, we used the integrand of the line-luminosity per volume, effectively the first moment of the line-luminosity function. The shot power at a given redshift is proportional to the second moment of the line-luminosity function (see Equation 8 in~\cite{COPSS} for the full relation including prefactors):
\be P_\text{shot}(z)\propto\int L^2\phi(L)\,d(\log{L})=\int L^2(M)\frac{dn}{d\log{M}}\,d(\log{M}).\ee We can calculate $L^2_\text{line}(M)\,dn/d(\log{M})$ using the mean $L_\text{line}(M)$ relations for our models with $dn/d(\log{M})$ from the same halo mass function fit from~\cite{Behroozi13b} that we used above for $dT/d(\log{M})$. We then re-express this as $dP_\text{shot}/dM$ to illustrate differential contributions to shot power from different halo mass ranges, in units of $\mu$K$^2$ Mpc$^3$. We show this in~\autoref{fig:dtdm} as well, along with the differential contributions to $\avg{T_\text{CO}}$ and $\avg{T_\text{HCN}}$.

} Breaking down the brightness temperature in this way, we can see that the strict cutoff at $10^{10}\,M_\odot$ likely affects $\avg{T_\text{CO}}$ significantly in our implementation of the power-law model\added{, but the effect is greatly diminished for $\avg{T_\text{HCN}}$ (as we anticipated in~\autoref{sec:hmll_mvar}) to the point of being negligible}. The CO temperature is also higher for the fiducial model because of the higher $L_\text{CO}(M)$ at $10^{12}\,M_\odot$ (as shown in~\autoref{fig:halol}), around the knee of the fiducial halo mass--CO luminosity relation. However, the power-law model predicts a large contribution to $\avg{T_{\rm HCN}}$ from high-mass emitters, which is not present in the \replaced{turnabout}{turnaround} model.\replaced{ We will revisit this class of HCN emitters in~\autoref{sec:res_halolum}, when we examine differences in the HCN luminosity function across these models.}{ We will return to these extreme emitters with halo mass $\gtrsim10^{13}\,M_\odot$ in the power-law model as we discuss the rest of our results, and specifically to the illustration of shot noise contributions by halo mass in~\autoref{sec:res_pspec}.}

Given the above, we may attribute a substantial part of the temperature differences between the fiducial model and the power-law models to the form of the halo mass--SFR relation. Accordingly, variations on the fiducial HCN model through tweaking $\alpha$ and $\beta$ in the $L_{\rm IR}$--$L'_{\rm HCN}$ relation do not result in differences as great. For $\alpha=1.09$ and $\beta=2.0$, following~\cite{Carilli05}, $\avg{T_{\rm HCN}}=0.083^{+0.012}_{-0.009}$ (median and 95\% sample interval). For $\alpha=1.23$ and $\beta=1.07$, following~\cite{GB12}, $\avg{T_{\rm HCN}}=0.052^{+0.007}_{-0.005}$ (median and 95\% sample interval).
\subsection{Power spectra}
\label{sec:res_pspec}

\begin{figure}
\centering\includegraphics[width=0.84\linewidth]{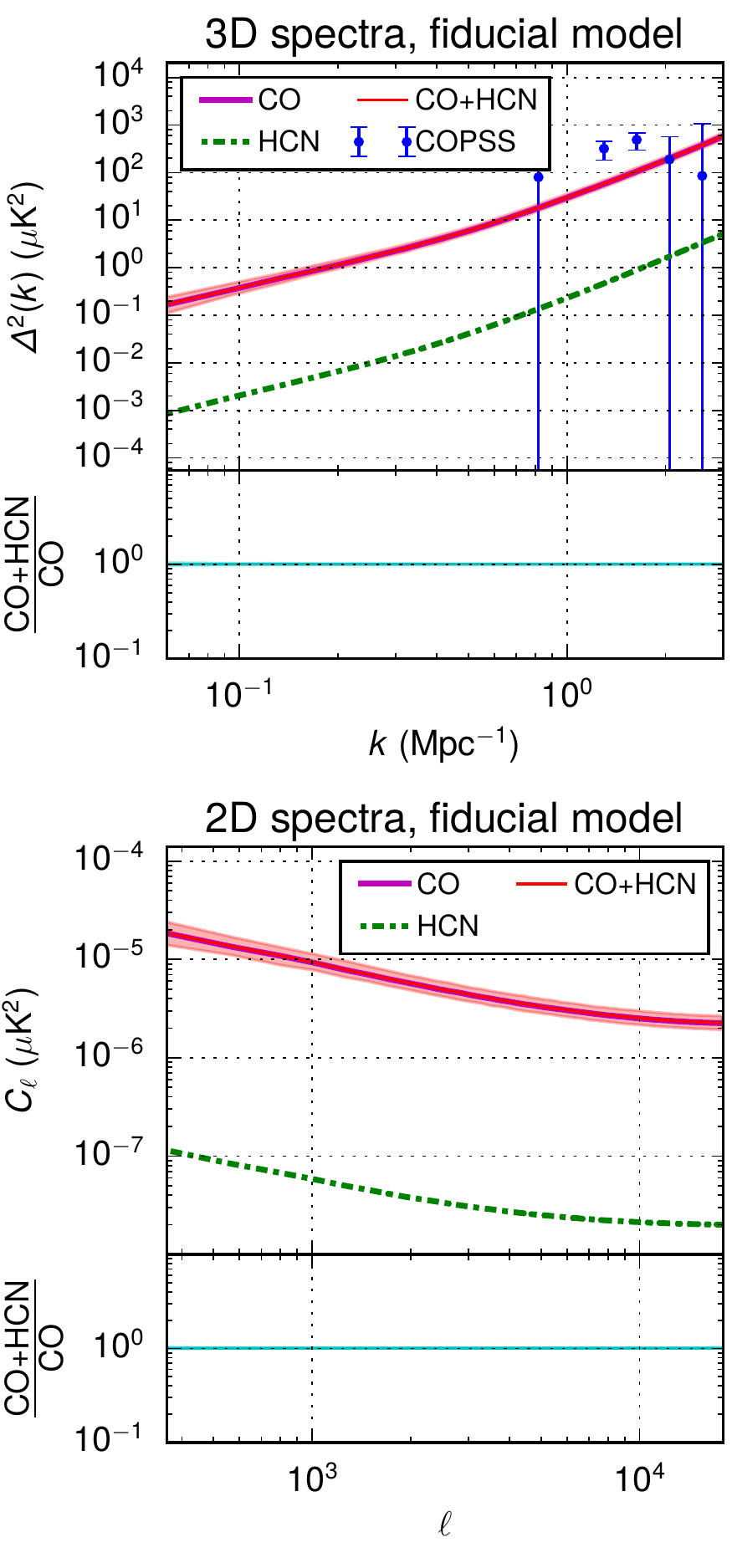}
\caption{3D spherically-averaged power spectra (\textbf{upper panel}) and average $C_\ell$ over all frequency channels (\textbf{lower panel}), for the fiducial model. Within each panel, the upper plot shows the auto spectrum for the CO signal only, the HCN contamination only, and the CO signal plus HCN contamination. Median spectrum values and 95\% sample interval (the latter shown only for CO plus HCN) at each $k$ or $\ell$ are shown for each model, with fractional residuals between uncontaminated and contaminated CO spectra shown below each spectra plot. We also show $\Delta^2(k)$ values from analysis of the full COPSS data set by~\cite{COPSS}.}
\label{fig:pspecfid}
\end{figure}

\begin{figure*}
\centering\includegraphics[width=0.942\linewidth]{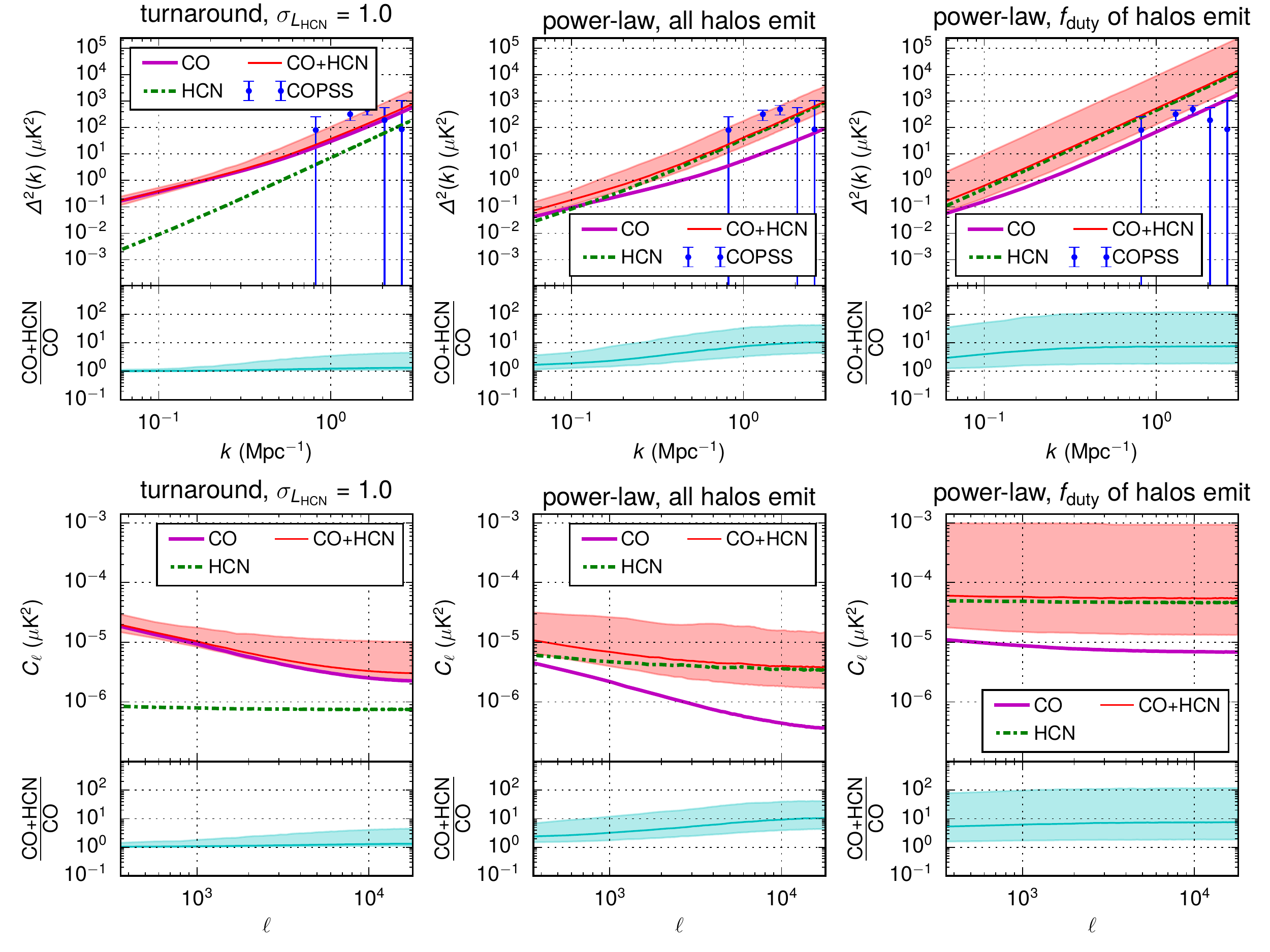}
\caption{3D spherically-averaged power spectra (\textbf{upper panels}) and average $C_\ell$ over all frequency channels (\textbf{lower panels}), for varied HCN line-luminosity models. Within each panel, the upper plot shows the auto spectrum for the CO signal only, the HCN contamination only, and the two combined. Median spectrum values and 95\% sample interval (the latter shown only for CO plus HCN) at each $k$ or $\ell$ are shown for each model, with fractional residuals between uncontaminated and contaminated CO spectra shown below each spectra plot. We again also show $\Delta^2(k)$ values from COPSS analysis by~\cite{COPSS}.}
\label{fig:pspec}
\end{figure*}

\autoref{fig:pspecfid} depicts the 3D and 2D power spectra from our simulated data cubes for the fiducial model. On top of median spectrum values, we also show the amount of lightcone-to-lightcone variation that exists in the CO spectra including contamination. 
For comparison, we also show $\Delta^2(k)$ values from COPSS as given in~\cite{COPSS}, which was published during the preparation of this work. These data constitute the first and presently only detection of the CO autocorrelation spectrum signal at any spatial mode as far as we are aware.

For our fiducial model, the HCN spectrum lies well below the CO spectrum, by a few orders of magnitude. However, the model variations from~\autoref{sec:hmll_mvar} affect the HCN spectrum to varying degrees, as~\autoref{fig:pspec} shows. Most notably, both of our power-law models place the HCN spectrum well above the CO spectrum, despite~\autoref{sec:res_halotemp} showing lower average cube temperatures in HCN than in CO for all models.\added{ The bias introduced in the total line intensity power spectra is similar between $\Delta^2(k)$ and $C_\ell$, although there is slightly less contamination in the former at the largest physical scales due to a stronger clustering signal. (This is to be expected---the survey volumes are elongated in the line-of-sight direction, so line-of-sight clustering ignored in the cube-averaged $C_\ell$ contributes significantly to $P(k)$ at lower $k$.)} Again, the power-law model has a different choice of IR--CO luminosity scaling that diminishes CO spectra in comparison to the fiducial spectra, but the relative rise in HCN spectra is due to no such thing. Spectrum contamination in the \replaced{turnabout}{turnaround} model is also possible if we increase log-scatter in HCN luminosity, although the median HCN spectrum even for $\sigma_{L_\text{HCN}}=1.0$ dex is not as high as the CO spectrum for smaller $k$ or $\ell$.

\added{Credible changes in the $L_{\rm IR}$--$L'_{\rm HCN}$ relation make little difference by comparison. Looking at the shot noise component of the power spectrum, the median $P_\text{HCN}(k=1\,\text{Mpc}^{-1})$ over all simulated observations goes up by a factor of 1.5 relative to the fiducial model if we use $\alpha=1.09$ and $\beta=2.0$, following~\cite{Carilli05}, and goes down by a factor of 2.2 relative to the fiducial model if we use $\alpha=1.23$ and $\beta=1.07$, following~\cite{GB12}. While these are relatively small changes in comparison to the orders-of-magnitude changes going between \replaced{turnabout}{turnaround} and power-law models, they do show that the level of empirical uncertainty in even the local IR--HCN connection alone implies a range of uncertainty of a factor of several in the HCN power spectrum.}

For both high log-scatter and power-law models, the HCN spectrum is very flat across the different modes (i.e.~$C_\ell$ and $P(k)$ are roughly constant, meaning $\Delta^2(k)\propto k^3$), and is entirely shot noise-dominated, unlike in the fiducial model.\added{ Again,~\autoref{fig:dtdm} analytically illustrates what bins of halo masses we expect to contribute most to this shot noise component of the signal. For the fiducial model, most of the shot noise for both CO and HCN comes from halos of mass $\sim10^{12}\,M_\odot$, at the $L_\text{line}(M)$ \replaced{turnabout}{downturn}. However, in the power-law model, the shot power originates from much higher-mass halos: the differential shot power peaks around $10^{13}\,M_\odot$ for CO and around $10^{14}\,M_\odot$ for HCN (which has a steeper IR--line luminosity relation than CO). So as in~\autoref{sec:res_halotemp}, we find a class of extreme HCN emitters only in the power-law models, with profound implications for predicted power spectra. We will revisit this class of emitters in~\autoref{sec:res_halolum}, where we examine differences in the numerically obtained HCN luminosity functions across these models.}


\subsection{Detection significance}
\label{sec:SNR}

\capstartfalse
\floattrue\begin{deluxetable}{ccc}
\tabletypesize{\footnotesize}
\tablewidth{0.9\linewidth}
\tablecaption{\label{tab:SNR}
Mean over all simulated observations of 496 lightcone pairs of total signal-to-noise ratio (SNR) for $P(k)$ over all modes.}
\tablehead{Model & CO (no HCN) & CO + HCN}
\startdata
\textbf{\replaced{turnabout}{turnaround}}:\\
$\sigma_{L_\text{HCN}}=0.0$ dex&2.19&2.20\\
{\bf $\sigma_{L_\text{HCN}}$ = 0.3 dex}&2.19&2.20\\
$\sigma_{L_\text{HCN}}=0.5$ dex&2.19&2.20\\
$\sigma_{L_\text{HCN}}=1.0$ dex&2.19&2.29\\
\hline
power-law:\\
$f_{\rm duty}$ of halos emit&1.17&6.97\\
all halos emit&0.64&1.38\\
\hline
\enddata
\tablecomments{The fiducial model parameters are indicated in bold. Models are as described in the main text in~\autoref{sec:hmll}, and in~\autoref{tab:meantemps}. SNR is given within the 30--34 GHz band, and per patch, not per survey.}
\end{deluxetable}
\capstarttrue

We use the CO detection significance metric used in \cite{Li15}, which is the sum in quadrature of the signal-to-noise ratio (SNR) of $P(k)$ over all $k$:
\begin{equation}\text{SNR}_\text{total}^2=\sum_i\left[\frac{P(k_i)}{\sigma_P(k_i)}\right]^2,\end{equation}
where $i$ indexes the binned spherical modes we obtain for $P(k)$. In their Appendix C,~\cite{Li15} outline the calculation of $\sigma_P(k)$, which incorporates Gaussian noise, sample variance from binning of the full $P(\mathbf{k})$ into shells, and resolution limits. (A beam size of $4'$ means that we effectively cannot meaningfully detect $P(k)$ beyond $k\sim1$ Mpc$^{-1}$, and for those modes $\sigma_P(k)\gg P(k)$. No beam smoothing is done in the cube, however.)

The extent to which HCN contamination \replaced{boosts}{biases} detection significance varies greatly, mostly depending on whether we use a power-law model or the \replaced{turnabout}{turnaround} model;~\autoref{tab:SNR} gives the mean total SNR over all modes for each model variation, and evidently the power-law models predict far higher \replaced{boosts in SNR from HCN contamination}{HCN contamination in total CO SNR}. While the CO spectra are comparatively lower in those models, this is not the sole reason the SNR \replaced{boost}{bias} is so much greater. If we mix models and use CO cubes generated with the fiducial model but HCN cubes generated with the power-law model (with all halos emitting), then there is still \replaced{a significant boost}{significant contamination} in SNR, \added{which rises} from 2.19 to 2.82---almost a 29\% increase compared to the $\lesssim1\%$ increase in the fiducial model. \added{We emphasize that these large increases are unlikely to be consistent with constraints on the galaxy--halo connection at high mass.}

As with mean map temperatures in~\autoref{sec:res_halotemp}\added{ and shot noise power spectra values in~\autoref{sec:res_pspec}}, varying the $L_{\rm IR}$--$L_{\rm HCN}'$ relation does not result in great differences, with the mean total SNR moving up from 2.19 to 2.21 following~\cite{Carilli05}, or 2.20 following~\cite{GB12}.

\cite{Li15} give signal-to-noise for a survey of four identical patches, not for a single patch. 
Accounting for this plus the increased instrument bandwidth and finer spectral resolution in the current COMAP design, we still expect the initial phase of COMAP to reach an overall detection significance near~\cite{Li15}'s estimate of approximately $8\sigma$.

\subsection{Luminosity functions}
\label{sec:res_halolum}

To inform discussion of the above results, we present HCN luminosity functions for our simulated emitters.

\begin{figure}\centering
\includegraphics[width=\linewidth]{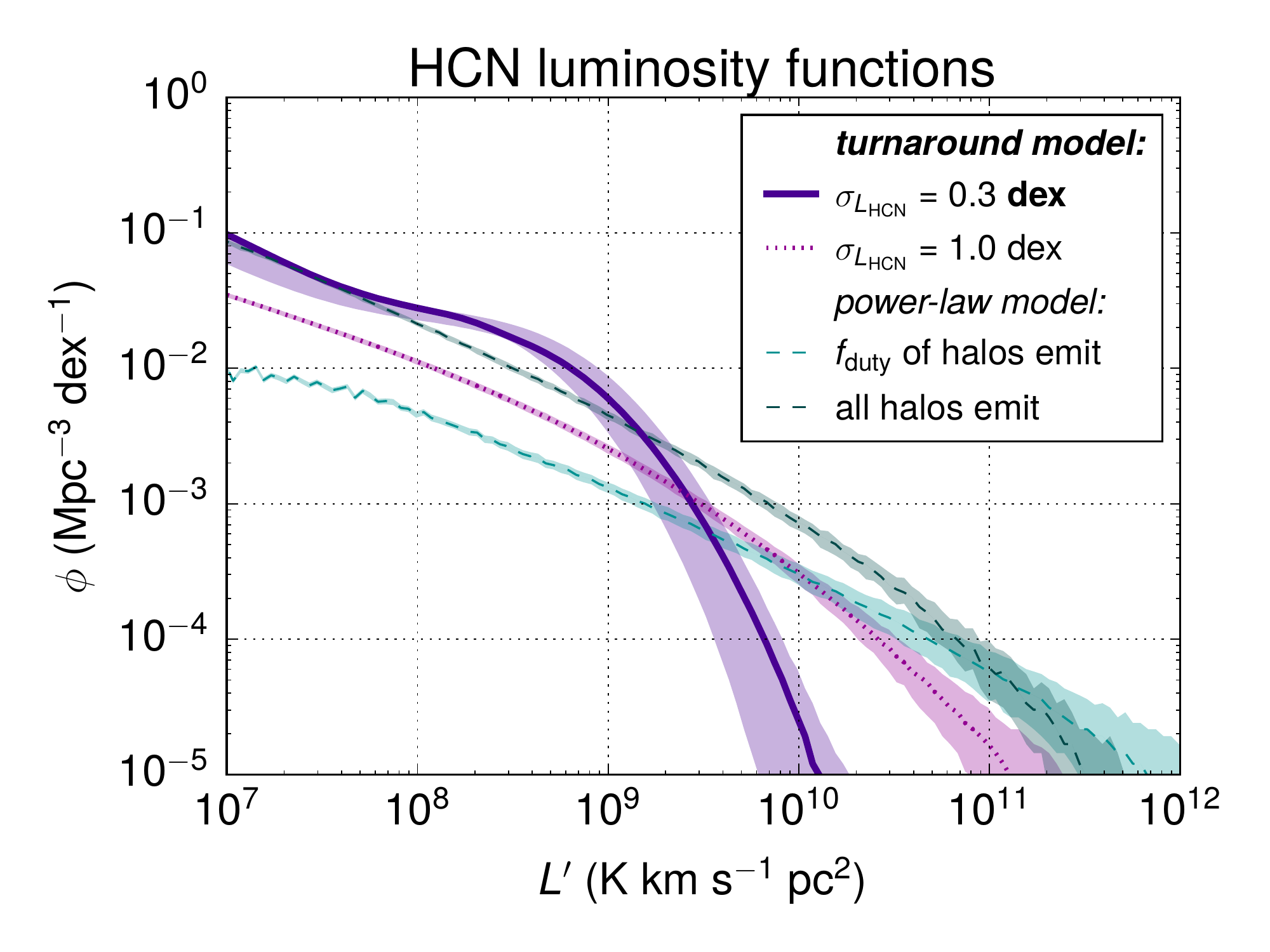}
\caption{Luminosity function $\phi(L)$ for HCN for varied line-luminosity models. The fiducial HCN model (the \replaced{turnabout}{turnaround} model with 0.3 dex scatter in line-luminosity) does not produce a significant population of halos above $L'\sim 10^9$ K km s$^{-1}$ pc$^2$, whereas the power-law models\added{ (plus the \replaced{turnabout}{turnaround} model with extremely broad 1.0 dex scatter in line-luminosity)} do produce a population of such bright (if still rare) emitters. Median values and 95\% sample intervals of $\phi(L)$ are depicted.}
\label{fig:model_halolum}
\end{figure}

\autoref{fig:model_halolum} shows the distribution of individual HCN luminosities as represented by the luminosity function $\phi(L)$, for a selection of models. (See~\autoref{sec:halolum_full} for details.) The power-law models result in luminosity functions that appear to approximately follow a single power law or at least a smooth function, whereas the fiducial model appears to exhibit an exponential cutoff beyond a `knee' of $L'\sim10^9$ K km s$^{-1}$ pc$^2$. Without a rapid cutoff in $\phi(L)$, the power-law models predict a significant population of extremely bright ($L'_{\rm HCN}\gtrsim10^{10}$ K km s$^{-1}$ pc$^2$), although still relatively rare, emitters---the same high-mass emitters shown in~\autoref{fig:dtdm} contributing a significant part of $\avg{T_\textnormal{HCN}}$.\footnote{The effect is not as drastic in CO, likely due to the higher $\alpha$ leading to lower-mass, lower-luminosity halos being more important in the first place. See~\autoref{fig:model_halolum_full} for an illustration.} The brightest HCN emitters would thus end up an order of magnitude brighter in our observations under those models than in our fiducial model. Variations on our fiducial model in $\alpha$ or $\beta$ in the IR--line-luminosity relation (as discussed in~\autoref{sec:hmll} and mentioned earlier in this section) do not alter luminosities or temperatures anywhere nearly as much.

Increased log-scatter in HCN luminosity in the \replaced{turnabout}{turnaround} model also results in broader distributions of luminosities and temperatures. With 1.0 dex scatter, the resulting $\phi(L)$ for HCN is qualitatively similar to what we obtain from the power-law models. \replaced{But broadness in $\phi(L)$ from halo-to-halo variance in luminosity is distinct from broadness that is simply due to the mean halo mass--line-luminosity relation, and we see the latter in the power-law models with or without implementing a duty cycle of $f_{\rm duty}\ll1$.}{The $\phi(L)$ curve takes on a broader shape simply due to the increased variance in HCN luminosity from halo to halo at fixed halo mass. However, in the case of the power-law models, the $\phi(L)$ curve takes on a broader shape even when the duty cycle is assumed to be unity and all halos emit, i.e.~without any halo-to-halo variance in HCN luminosity for a given halo mass. Therefore, the luminosity function varies by model both due to different halo-to-halo scatter in luminosity and due to the basic form of the mean halo mass--line-luminosity relation.

}Note also that 1.0 dex scatter in the relation is so high that the correlations that our references claim would scarcely be observable. Even at high redshift, we have no reason to believe there would not be such an observable correlation.
\section{Discussion}
\label{sec:discuss}
\subsection{Comparison with previous work}
\cite{BKK15} provide our main point of comparison for line contamination forecasts, with simulated 2D maps of emission (for a single observing frequency) in a variety of lines, including CO and HCN.~That work saw large simulation uncertainty due to variations in halo numbers and luminosities, but generally saw HCN spectra to be within an order of magnitude of CO spectra and found it to significantly affect observed power spectra, although no quantitative measure of this effect is provided. Since that work used 2D maps (550 deg$^2$ each), it uses $C_\ell$ to characterise the auto spectra in these maps at different scales, with $\ell$ ranging from $10^1$ to $\gtrsim10^3$. HCN amplitude begins to even exceed CO amplitude beyond $\ell\sim10^3$.

Early single-dish surveys like COMAP, due to their small survey field size, will probe a range of $\ell$ about an order of magnitude higher than the mock surveys of~\cite{BKK15}, at $\sim10^2$ to $\sim10^4$. The sensitivity of COPSS peaks at the upper end of this range. Thus, the effect of HCN contamination predicted by the power-law models, meant to mimic the model of~\cite{BKK15}, is even more striking. Yet in the results obtained with our fiducial \replaced{turnabout}{turnaround} model, the effect appears negligible, with the HCN auto spectra lying several orders of magnitude below CO spectra for all spatial modes, in both 2D and 3D data.

\subsubsection{Lessons from model variations}

Two factors influence the level of HCN contamination in our simulated observations.
\begin{enumerate}
\item Does the model incorporate the commonly expected fall-off in star-formation efficiency beyond halo masses of $\sim10^{12}$ $M_\odot$? If not---and if SFR, IR luminosity, and line-luminosities remain well-correlated at high mass and high redshift---a population of very bright, very sparsely distributed HCN emitters from intermediate redshift will overshadow the high-redshift CO signal.
\item How much stochasticity or scatter is in the halo mass--line-luminosity relation? If the galaxy population at intermediate redshift ends up with far more dramatic gas dynamics and time-evolution characteristics as a whole, we may expect extreme intrinsic scatter in line-luminosity in comparison to the high-redshift population. The resulting amount of scatter and thus shot noise in the HCN contamination could overpower the CO spectrum. However, this effect is subject to lightcone-to-lightcone sample variance, and not necessarily alarming in the context of an initial overall CO $P(k)$ detection at the scales that COMAP targets.
\end{enumerate}

The first point is particularly relevant, as past work on line contaminants and often on line-intensity surveys in general---including~\cite{Pullen13} and~\cite{BKK15}---have considered simpler linear or power-law halo mass--line-luminosity relations, much like the power-law models in this work. Some works like~\cite{Gong11} and~\cite{Li15} do incorporate a non-monotonic relation not described by a simple single power law. In fact, previous work supports a non-power-law relation between halo mass and SFR, as~\cite{Behroozi13a} note. Compared to the power-law models, the fiducial model also predicts lower HCN line-luminosities that tie more closely with HCN detections and upper limits at high redshift, as we discuss briefly in~\autoref{sec:obsHCN}.

The shape of the halo mass--HCN luminosity relation significantly affects the contribution of the high-mass halo population in simulated observations, and thus the conclusions drawn from those simulations. Predicted HCN spectra are higher with the power-law models compared to the fiducial model by several orders of magnitude, and when predicting \replaced{boosts}{bias} in CO detection significance due to HCN contamination, power-law model predictions far exceed the fiducial predictions. As the luminosity functions in~\autoref{sec:res_halolum} demonstrate, the power-law models result in more high-luminosity halos, which are still rare enough that their spatial distribution within each lightcone is quite sparse (see~\autoref{fig:slice_viz} for a visual reference), and their statistics vary greatly from lightcone to lightcone. Thus, both 2D and 3D power spectra take on a flat shape and a wide 95\% sample interval across repeated simulations, most of the spread being due to HCN rather than CO.

Note that the second factor---the amount of stochasticity or scatter present---includes the exact implementation of $f_\text{duty}$ in the power-law models. Introducing a duty cycle of $f_\text{duty}\ll1$, in particular, leads to extreme halo-to-halo scatter in line-luminosity, and thus to shot noise dominating the power spectrum for all molecular lines. However, we see similar differences between the fiducial models and the power-law models, with or without selecting only $f_\text{duty}$ of halos---so $f_\text{duty}$ is not solely responsible for the differences that we see.

\subsubsection{Discrepancies against~\cite{BKK15}}
\label{sec:bkk15_comp}
The results in this work and the results given by~\cite{BKK15} diverge quantitatively, even with the same halo mass--line-luminosity models. Simulated observations generated via the power-law models have $\avg{T_{\rm CO}}\approx0.47$ $\mu$K and $\avg{T_{\rm HCN}}\approx0.14$ $\mu$K, compared to predictions in~\cite{BKK15} of $\avg{T_{\rm CO}}\approx0.60$ $\mu$K and $\avg{T_{\rm HCN}}\approx0.023$ $\mu$K. Values of $C_\ell$ are also somewhat elevated relative to~\cite{BKK15}, at least at $\ell\sim10^3$, where our range of $\ell$ overlaps with the range of $\ell$ in~\cite{BKK15}. Comparing between~\autoref{fig:pspec} in this work and Figure 1 in~\cite{BKK15}, the contrast is more drastic for HCN than for CO.

We are still exploring reasons behind these discrepancies and what factors affect them. However, they likely arise from differences between the two works in halo mass cutoff, halo mass completeness, and halo mass function. We noted the first two of these in~\autoref{sec:hmll_mvar}. Both are critical for an accurate simulation of CO emission, since lower-mass emitters can contribute non-negligibly to the CO signal. As we can see in~\autoref{fig:dtdm}, our simulations lack some of these fainter emitters due to the strict mass cutoff of $10^{10}\,M_\odot$, and this contributes to the lower sky-averaged CO temperature in comparison to~\cite{BKK15}. Note also that the fiducial CO spectrum is somewhat lower than the COPSS measurements from~\cite{COPSS}, which we could ascribe to missing faint CO emitters in our simulations.

Furthermore, even minor differences in the halo mass function will affect the power spectrum noticeably---particularly for HCN, as the choice of mass function in the approach of~\cite{BKK15} would significantly impact the high-mass halo population. As discussed above, this population is small but a dominant influence in the HCN auto spectrum.


Broadly, our results for both of the power-law models are in line with~\cite{BKK15}: \emph{if} we assume simple power-law halo mass--line-luminosity relations, we expect that the brightest HCN emitters raise HCN spectrum amplitudes to be on par with CO spectrum amplitudes on the spatial scales being observed, and expect significant contamination of CO spectra. However, we do not have high confidence in the power-law relations underlying these models, as discussed above.


\subsection{Implications for CO surveys}

For COMAP and other near-future surveys with a goal of initial CO signal detection, we find HCN contamination may not pose the most significant risk to such an initial detection. Under our fiducial model, the effect on total detection significance is sub-percent level. Given the uncertainty in relative intensities of CO and HCN, and the lack of data on the HCN luminosity function, we should not completely dismiss HCN as a possible contaminant. However, mitigation of HCN should be placed at a lower priority than mitigation of other systematics and astrophysical foregrounds.

The net effect of line contamination may be several times higher than forecast here---perhaps 5 to 10 percent in terms of relative \replaced{boost}{bias} in total CO $P(k)$ detection significance---once we incorporate consistent modelling of other lines, many of which we expect to have average luminosities similar to HCN.~\cite{BKK15} consider a variety of line contaminants beyond HCN, including CN(1-0) and HCO+(1-0), emitting respectively at rest frequencies of $113$ GHz and $89$ GHz, close to CO(1-0) and HCN(1-0).\footnote{We expect HCO+ in particular to emit at luminosities similar to HCN, contrary to estimates shown in~\cite{BKK15} at the time of writing.}

However, all of these contaminant lines trace denser gas than CO, and emit at lower rest frequencies than CO(1-0). Both facts work to our advantage in a CO intensity mapping survey: much of our signal will come from aggregated fainter galaxies, where lower gas densities and the frequency-dependence of thermal emission enhance CO emission above denser tracers.

For far-future surveys at similar or higher redshifts, which may attempt to extract more sophisticated information about galaxy assembly and star-formation, the impact of HCN contamination could potentially be more marked. Higher angular resolution and instrumental sensitivity would ideally allow us to detect higher-$k$ modes. However, the effect of HCN contamination in $\Delta^2(k)$ only worsens for higher-$k$ modes in our simulated observations, for all models considered. \cite{BKK15b} describe how such contamination affects inferences about the galaxy population based on voxel intensity distributions. For the more pessimistic models of HCN emission, we may expect similarly adverse effects on the inferences that~\cite{Li15} present, given the contamination we see in $\Delta^2(k)$. 

Many future CO(1-0) intensity surveys will cross-correlate against other large-scale structure tracers, potentially including higher-order CO lines. While such analysis would exclude systematics and foregrounds not common to both data sets, they would not exclude molecular line foregrounds like HCN with their own higher-order lines. HCN(2-1)--HCN(1-0) contamination in a CO(2-1)--CO(1-0) survey would likely be of the same level of concern as HCN(1-0) in COMAP, since we expect the 2-1 and 1-0 lines to be of similar luminosity. (In their analysis of HCN detections, \cite{GB12} adopt a line ratio of $L'_{\rm HCN(2\operatorname{-}1)}/L'_{\rm HCN(1\operatorname{-}0)}=0.7$ from an empirical mean.) To best characterise the target and contaminant signals in isolation, we must cross-correlate against large-scale structure tracers beyond line-intensity surveys, such as QSO or galaxy surveys as explored in~\cite{Pullen13}. \added{As that paper suggests, a cross-correlation signal between such tracers and any molecular line---including HCN---would be of interest in its own right, although outside the scope of this study.} 

In the interim, our predictions for both target and contaminant signals will change, for better or worse. The results presented above use empirical models that build largely on observations of bright, local galaxies. Furthermore, these models build on a connection between star-formation activity and IR or line-luminosities, which is not entirely certain. Some or all of our current modelling may thus extrapolate poorly to faint galaxies or dwarf galaxies, and we must eventually move to physically motivated models of line-luminosity or spectral emission density templates to simulate line emission at high redshift in multiple lines simultaneously.

Changes in these predictions may come rapidly with advances in understanding of star-formation activity and its relation to molecular and dense gas. Such advances will arise not only from line-intensity surveys like COMAP, but also further observations of individual galaxies and clusters through ALMA and VLA, as well as the more recently commissioned {\it Argus} spectrometer at the Green Bank Observatory~\citep{Argus}.

\section{Conclusions}
\label{sec:conclude}

Our \replaced{simulations indicate the following findings}{primary conclusions are as follows}:
\begin{itemize}
\item Under a basic power-law model, simulated HCN emission potentially seriously affects our CO power spectra and detection significance.
\item However, in \replaced{the fiducial model}{a more realistic model based on knowledge of high mass galaxies}, we find that simulated HCN emission in a CO survey lies an order of magnitude below CO emission in temperature.
\item The HCN auto spectrum also lies several orders of magnitude below the CO auto spectrum, across all spatial modes. The resulting \replaced{boost}{contamination} in total detection significance is a small effect given reasonable amounts of halo-to-halo scatter.
\end{itemize}

Our fiducial model is somewhat better motivated than previous models, and predicts no serious contamination of CO power spectra. Still, we do caution restraint in dismissing HCN as a contaminant, given the limited observational information we have for high-redshift HCN(1-0) sources, and the plausibility of high intrinsic scatter at these redshifts. An investigation of high-redshift, high-mass halos hosting higher-luminosity ($L'_{\rm HCN}\gtrsim10^{10}$ K km s$^{-1}$ pc$^2$) HCN emitters would help constrain the duty cycle and luminosity scaling for HCN emission, which would refine future modelling.

In the run-up to surveys like COMAP, we expect further work on line contaminants, especially on the effect of such contamination on cross-correlation between complementary surveys, and the development of more sophisticated models and mitigation strategies for these contaminants. Indeed, during the preparation of this work,~\cite{Lidz16} and~\cite{Cheng16}, working in the context of [CII] intensity surveys, have shown the possibility of separating strong CO line contamination from the targeted [CII] emission (at least at the power spectrum level), and even extracting useful information about both target and contaminant lines. While our expectation is that HCN contamination does not pose the greatest risk to CO intensity mapping, such techniques would readily find use in CO surveys if that expectation should change.


\acknowledgments{This work was supported by NSF AST-1517598 and by a seed grant from the Kavli Institute for Particle Astrophysics and Cosmology. We thank Patrick Breysse, Kieran Cleary, Andrew Harris, Brandon Hensley, and members of the COMAP collaboration\added{, as well as an anonymous referee,} for helpful discussions and comments. We thank Matthew Becker for providing access to the Chinchilla cosmological simulation (\texttt{c400-2048}) used in this work. This research made use of Astropy, a community-developed core Python package for astronomy~\citep{astropy}, as well as\added{ Matplotlib~\citep{matplotlib} and} a modified version of \texttt{hmf}~\citep{hmf}.\added{ This research made use of NASA's Astrophysics Data System Bibliographic Services.} This work used computational resources at the SLAC National Accelerator Laboratory.}


\bibliographystyle{aasjournal}
\bibliography{references}

\appendix
\section{Observational checks on simulated line-luminosity}
\label{sec:obsHCN}
\cite{Li15} provide a brief comparison between the CO emission model in this work and observed CO luminosities, which showed general consistency. We attempt a similar comparison for our HCN models, but the sample of high-redshift HCN-emitting galaxies that we can use is much more limited, in number and in inferred properties. Unlike the references in~\cite{Li15}, our references give no stellar masses for the galaxies observed, only gas masses and dust masses for some. Since we have no direct relation between halo mass and either of those masses, we will attempt only a very simple sanity check between our models and the observations available in the literature.

\capstartfalse\floattable
\begin{deluxetable}{cccccc}[t]
\tabletypesize{\footnotesize}
\tablecaption{\label{tab:obsHCN}
Sample of high-redshift galaxies with HCN(1-0) detections or upper limits.}
\tablehead{
\colhead{Source} & \colhead{$z$} & \colhead{lens mag.} & \colhead{$L_\text{IR}$} & \colhead{$L'_\text{CO}$} & \colhead{$L'_\text{HCN}$}\\
&&&($10^{12}$ $L_\odot$)&\multicolumn{2}{c}{($10^9$ K km s$^{-1}$ pc$^2$)}}
\startdata
VCV J1409+5628&2.583&1&17&74&6.5\\
APM 08279+5255&3.911&80&0.25&0.92&0.25\\
H1413/Cloverleaf&2.558&11&5.0&37&3.0\\
IRAS F10214+4724&2.286&17&3.4&6.5&1.2\\
J16359+6612(B)&2.517&22&0.93&3.7&0.6\\
BR 1202-0725&4.694&1&55&93&$<39.0$\\
SMM J04135+1027&2.846&1.3&22&159&$<28.0$\\
SMM J02399-0136&2.808&2.5&28&112&$<46.0$\\
SDSS J1148+5251&6.419&1&20&25&$<9.3$\\
SMM J02396-0134&1.062&2.5&6.1&19&$<3.7$\\
SMM J14011+0252&2.565&25--5&0.7--3.7&4--18&$<0.3$--1.5\\
MG0751+2716&3.200&17&2.7&9.3&$<0.9$\\
RX J0911+0551&2.796&22&2.1&4.8&$<0.6$\tabularnewline
\enddata
\tablecomments{Redshifts compiled in~\cite{SVB05} and luminosities compiled in~\cite{GS07}. All quantities are intrinsic, not apparent, and account for lens magnification as assumed in~\cite{GS07}. For SMM J14011+0252 we use the highest quoted values (assuming lowest lens magnification) for each luminosity in~\autoref{fig:obsHCN}.}
\end{deluxetable}
\capstarttrue

\autoref{tab:obsHCN} shows a selection of high-redshift sources, as compiled in~\cite{SVB05} and~\cite{CW13}, for which~\cite{GS07} (and references in each) give luminosities in IR, CO(1-0), and HCN(1-0). To compare these properties with what we expect from our models, we convert halo mass to the same luminosities as in~\autoref{sec:hmll}, incorporating log-scatter in each relation. We consider a halo mass of $10^{13}$ $M_\odot$, which is a probable halo mass size for LIRGs~\citep{LIRGHM}, and also use a lower halo mass of $10^{12}$ $M_\odot$. We then look at the scatter of the resulting sample predictions for both $L_\text{HCN}'$ and $L_\text{IR}$ or $L'_\text{CO}$, for a selection of high redshifts.

\begin{figure*}
\begin{center}\includegraphics[width=0.96\linewidth]{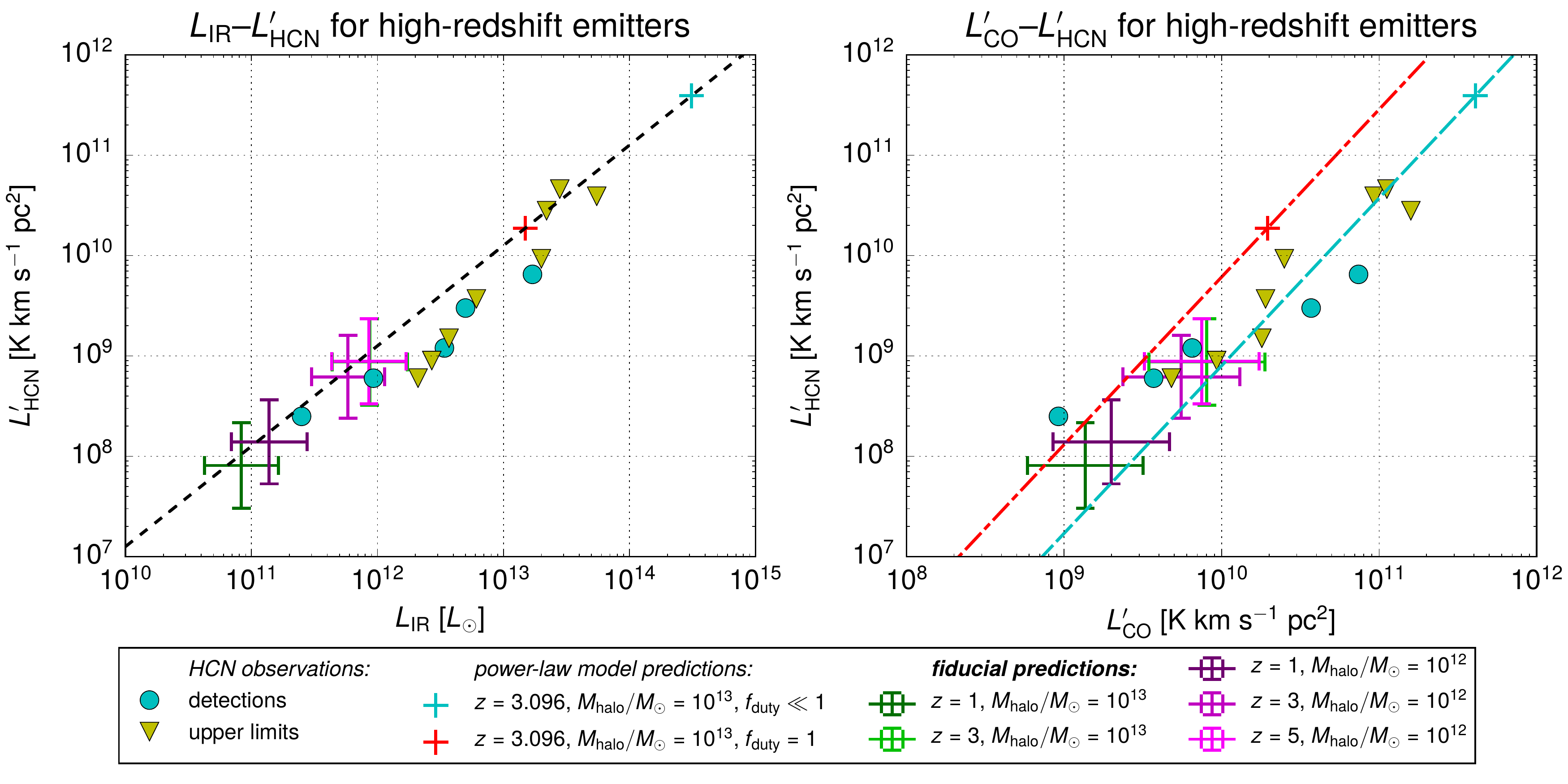}\end{center}
\caption{Predicted HCN line-luminosity against IR luminosity ({\bf left}) and CO(1-0) luminosity ({\bf right}), for a range of redshifts and halo masses (indicated in the plot legend). Error bars\added{ on fiducial predictions} show marginal $1\sigma$ intervals for each luminosity. We base all luminosities on the fiducial models of~\autoref{sec:hmll}. We over-plot observed HCN luminosities from~\autoref{tab:obsHCN} (cyan circles for detections, yellow triangles for upper limits). \replaced{We also over-plot predictions from a power-law model with $f_{\rm duty}$ included in the relation (dash-dotted line, magenta) and without it (dashed line, green), for a fixed redshift $z=3.096$ (the mean redshift of the observed high-redshift sample) and a fixed halo mass $M_\text{halo}=10^{13}$ $M_\odot$. The magenta dash-dotted line also doubles as a prediction for a power-law model with $f_{\rm duty}$ excluded from the relation, but at the same redshift and a halo mass closer to $1.6\times10^{12}$ $M_\odot$.}{On the left panel, we over-plot the $L_\text{IR}$--$L'_\text{HCN}$ fit from~\cite{GS04} (dashed line, black), used in both the power-law and \replaced{turnabout}{turnaround} models. On the right panel, we also plot the CO--HCN luminosity relation for the power-law model, both with $f_{\rm duty}\ll1$ absorbed into the luminosities and assuming a duty cycle of unity (dash-dotted line, red), and without it absorbed into the luminosities and assuming a duty cycle of $f_{\rm duty}$ (dashed line, cyan). We plot those relations only for $z=3.096$, the mean redshift of the observed high-redshift sample.}}
\label{fig:obsHCN}
\end{figure*}

\autoref{fig:obsHCN} shows sample predictions over-plotted with the observations. We find that the HCN detections are broadly consistent with our\deleted{ fiducial }model predictions\replaced{, and more in line with the fiducial model than with the power-law models.}{, although the power-law model prediction is arguably overly bright if assuming a duty cycle of $f_\text{duty}\ll1$ (see the cyan plus marker in the plots compared to the observed detections and upper limits, as well as the distributions of the fiducial predictions). Many detections are brighter in IR or line emission than the fiducial model predicts, but the fiducial model aims to describe relatively normal early-universe galaxies. By contrast, these detections are not only lensed (with magnifications of unspecified uncertainty), but also most analogous to extreme starbursts or ULIRGs, as~\cite{GS07} note. It may be entirely reasonable to expect that in certain environments, as those experienced by starburst galaxies, the fall-off in star-formation efficiency may be slower or occur at higher mass than in other environments. Nonetheless, even though our models gloss over such factors in their simplicity, the predicted HCN luminosities for the input halo masses are essentially sane.}

Note that the halo mass is not directly observable for any of the high-redshift galaxies in~\autoref{tab:obsHCN}, and we have assumed plausible halo masses for the purpose of this sanity check purely on theoretical grounds, without any firm observational basis. Adjusting these halo masses up or down by one or two orders of magnitude would drastically change our models' luminosities. Thus we do not propose to either confirm or rule out any models with this sanity check, which merely shows that none of the models in this work assign outlandish luminosities to CO and HCN emitters. Additionally, we remind ourselves again that our models and more generally any of these ladders of observation- and simulation-based empirical relations are a decidedly primitive description of radio, sub-mm, and IR emission from dark matter halos.

We note the apparent scarcity of observations of HCN(1-0) emission in high redshift galaxies. The supplementary information for~\cite{CW13} lists 61 detections of CO(1-0), but only 3 detections of HCN(1-0). These three were in the Cloverleaf, IRAS F10214+4724, and VCV J1409+5628 QSOs, with all three detections already compiled in~\cite{GS07}. Curiously,~\cite{CW13} also lists no CO(1-0) detection for VCV J1409+5628, so we should not view its compilation of detections in the literature as complete by any means.
\section{Calculation of luminosity functions}
\label{sec:halolum_full}
\begin{figure}\begin{center}
\includegraphics[width=0.42\linewidth]{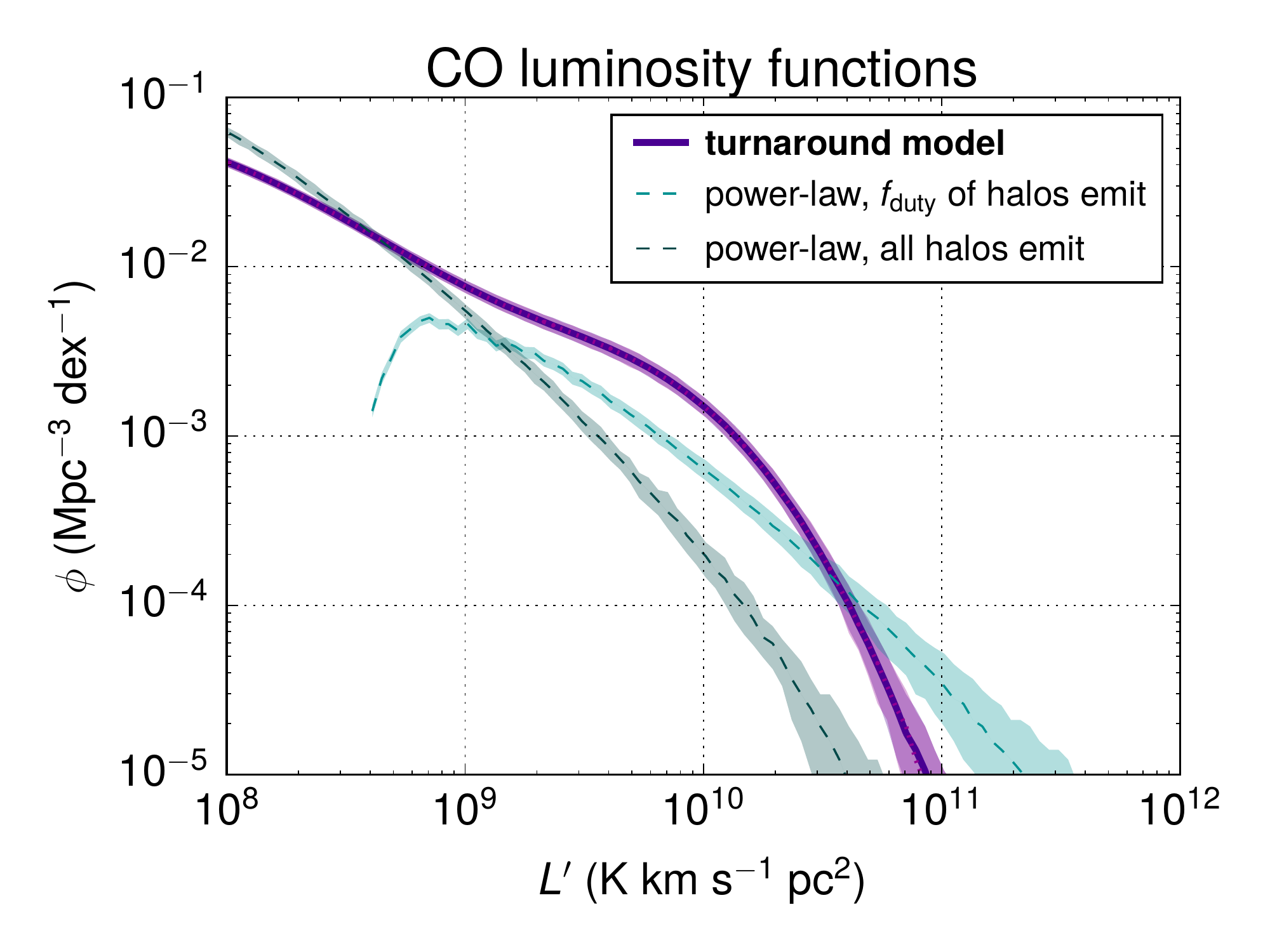}\quad\includegraphics[width=0.42\linewidth]{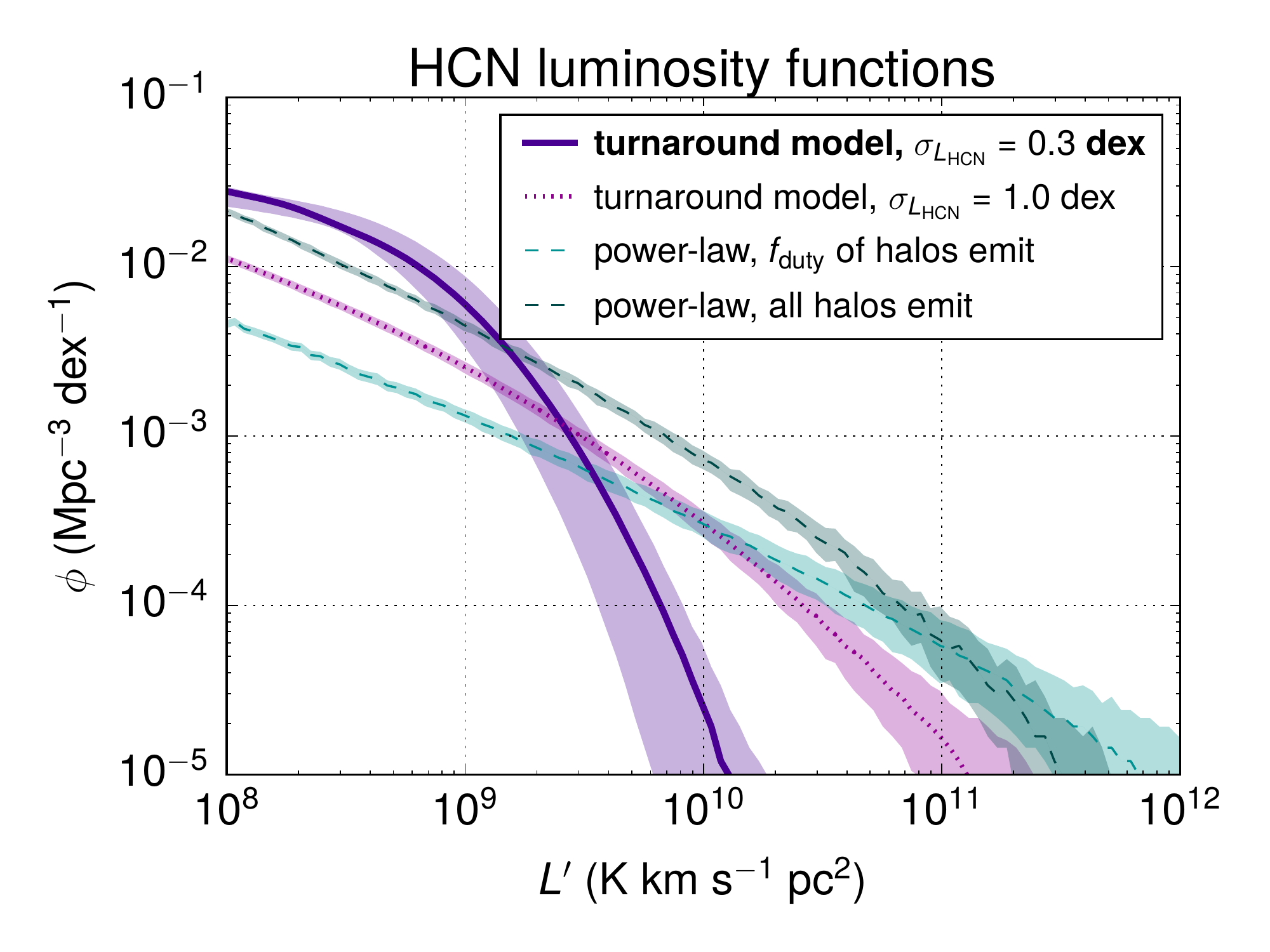}\end{center}
\caption{Luminosity function $\phi(L)$ for CO (\textbf{left panel}) and HCN (\textbf{right panel}) for varied line-luminosity models. The fiducial CO and HCN models carry 0.3 dex scatter in line-luminosity.\added{ For the \replaced{turnabout}{turnaround} model with $\sigma_{L_\text{HCN}}=1.0$ dex, the HCN luminosity function is shown separately from the fiducial model's $\phi(L)$, but the change in HCN luminosity scatter does not (and should not) affect the CO luminosity function.} Median values and 95\% sample intervals of $\phi(L)$ are depicted.}
\label{fig:model_halolum_full}
\end{figure}

As an extension of the work in~\autoref{sec:res_halolum}, we present in~\autoref{fig:model_halolum_full} both CO and HCN luminosity functions for individual halos (only down to $L'\sim10^8$ K km s$^{-1}$ pc$^2$, to match results presented for CO(1-0) by~\cite{Li15},~\cite{COPSS}, and others). We take counts of halos $N(L)$ in log-space luminosity bins $(L,10^{\Delta(\log{L})}L)$, and calculate the luminosity function as
\begin{equation}\phi(L) = \frac{N(L)}{\Delta(\log{L})\,\Delta V},\end{equation}
where $\Delta V$ is the lightcone comoving volume. Thus $\phi$ is number volume density per luminosity bin width, which is the typical definition of the luminosity function. Note the falloff and ringing of $\phi$ at extreme luminosity values (most visible for the power-law model with only $f_{\rm duty}$ of halos emitting), which is at least in part an artefact of the parameters of the simulation (the cutoff mass of $10^{10}\ M_\odot$ combined with the relatively high particle mass of the cosmological simulation).

Otherwise, the values for $\phi$ over the given range are not unreasonable for any of our models. While no space density data for HCN emitters appear to be available, the simulated CO luminosity function values for the models in this work do not compare unfavourably to data and fits that~\cite{COLF} and~\cite{COLFALMA} present in the range of $L'\gtrsim 10^9$ K km s$^{-1}$ pc$^2$, where CO observational data are chiefly available. The overall features of the CO luminosity function, including the knee at $L'_{\rm CO}\sim10^{10}$ K km s$^{-1}$ pc$^2$, are entirely consistent with the $1\sigma$ constraints from COPSS data given in~\cite{COPSS}.

\listofchanges
\end{document}